\documentclass[twocolumn]{aastex6}

\usepackage{mathptmx}
\usepackage[T1]{fontenc}
\usepackage{ae,aecompl}
\usepackage{graphicx}	
\usepackage{amsmath}	
\usepackage{amssymb}
\usepackage{tabularx}
\usepackage{bm}
\usepackage{morefloats}

\def\feh{\mathrm{[Fe/H]}}
\def\afe{[\alpha/\mathrm{Fe}]}

\def\vlos{v_\mathrm{los}}
\def\disp{\sigma_\mathrm{int}}

\def\scon{\sigma_\mathrm{contam}}

\shorttitle{Substructure in A267}
\shortauthors{Tucker et. al.}

\begin{document}
\title{Galaxy Cluster Mass Estimates in the Presence of Substructure}

\author{Evan Tucker}
\author{Matthew G. Walker}
\affil{McWilliams Center for Cosmology, Carnegie Mellon University, Pittsburgh, PA, USA}
\author{Mario Mateo}
\affil{Department of Astronomy, University of Michigan, Ann Arbor, MI 48109-1042}
\author{Edward W. Olszewski}
\affil{Steward Observatory, University of Arizona, Tucson, AZ 85721}
\author{Alex Geringer-Sameth}
\affil{Department of Physics, Imperial College London, Prince Consort Rd, London SW7 2AZ, UK}
\and
\author{Christopher J. Miller}
\affil{Department of Astronomy, University of Michigan, Ann Arbor, MI 48109, USA}

% Abstract of the paper
\begin{abstract}
We develop and implement a model to analyze the internal kinematics of galaxy clusters that may contain subpopulations of galaxies that do not independently trace the cluster potential.  The model allows for substructures within the cluster environment, disentangles cluster members from contaminating foreground and background galaxies, and includes an overall cluster rotation term as part of the cluster kinematics.
We estimate the cluster velocity dispersion and/or mass while marginalizing over uncertainties in all of the above complexities.  In a first application to our published data for Abell 267 (A267), we find no evidence for cluster rotation but we identify up to five distinct galaxy subpopulations.  We use these results to explore the sensitivity of inferred cluster properties to the treatment of substructure.  Compared to a model that assumes no substructure, our substructure model reduces the dynamical mass of A267 by $\sim 20\%$ and shifts the cluster mean velocity by $\sim 100$ km s$^{-1}$, approximately doubling the offset with respect to the velocity of A267's brightest cluster galaxy.  
Embedding the spherical Jeans equation within this framework, we infer for A267 a dark matter halo of mass $M_{200}=6.77\pm1.06\times10^{14}M_\odot/h$, concentration $\log_{10}c_{200}=0.61\pm0.39$, consistent with the mass-concentration relation found in cosmological simulations. 
\end{abstract}


%%%%%%%%%%%%%%%%%%%%%%%%%%%%%%%%%%%%%%%%%%%%%%%%%%

%%%%%%%%%%%%%%%%% BODY OF PAPER %%%%%%%%%%%%%%%%%%

\section{Introduction}
\label{intro}

Galaxy clusters are the most massive gravitationally bound and relaxed structures in the Universe, thereby representing important laboratories for observational cosmology \citep[][]{Rines2003, Voit2005, Jones2009, Vikhlinin2009, Geller2013, Rines2013, Sohn2017}.
Due to their high density of galaxies they are also ideal for studying galaxy interactions and the effect these interactions have on the galaxy population.
Galaxy clusters are studied in a multitude of ways, from gravitational lensing both weak and strong \citep[for example][and references therein]{Kneib2008, Postman2012, Applegate2014, Barreira2015, Gonzalez2015}, to X-ray temperature measurements of hot intracluster gas \citep[][]{Guennou2014, Moffat2014, Girardi2016, Rabitz2017} to Sunyaev-Zeldovich effects \citep[][]{Sunyaev1970, Churazov2015}, to spectroscopic velocity measurements of cluster members \citep[e.g.][and references therein]{Rines2003, Geller2014, Stock2015, Rines2016, Tasca2016, Biviano2016}.
All of these methods can provide mass estimates, thus constraining the high-mass end of the halo mass function, thereby constraining cosmological parameters such as the amplitude of the power spectrum or the evolution of dark matter and dark energy density parameters.

When calculating cluster masses using the velocities of cluster members, it is common to assume that the cluster is a relaxed system with a gravitational potential and kinematics that satisfy the viral theorem.
However, such assumptions neglect recent galaxy accretion that could alter the distribution of galaxies in phase space \citep[][]{Regos1989, vanHaarlem1993, Diaferio1997, Rines2003}.
Residual angular momentum during formation, as well as the presence of in-falling groups, could contribute a rotational velocity to the cluster \citep[][]{Aryal2013, Tovmassian2015, Manolopoulou2017}.
\citet{Li1998} suggests that any global rotation of the universe could provide angular momentum to galaxy clusters during their formation.
Additionally, even in systems that appear relaxed, these mergers can generate residual substructure within the cluster environment such that individual galaxies are not necessarily independent tracers of the gravitational potential \citep[][]{Dressler1988, Biviano2002, Girardi2015}.
These factors have the potential to impact dynamical mass measurements, leading to systematic errors which will then propagate into cosmological inferences.

Early efforts were made to detect rotations in galaxy clusters; however, this proved difficult without distinguishing between closely interacting groups \citep[see][for example]{Materne1983, Oegerle1992}.
More recently the effects of recent mergers and close interactions have been accounted for and some authors have started exploring galaxy rotation in more depth.
Some have used large surveys such as SDSS to look for galaxy rotation in relaxed systems and report evidence of rotating clusters \citep[][]{Hwang2007, Tovmassian2015}.
Multiple analyses of Abell 2107 have concluded that it is rotating \citep[][]{Oegerle1992, Kalinkov2005}.
Through X-ray observations some groups have studied the rotation of the intracluster medium (ICM) \citep[e.g.][]{Bianconi2013}.
And most recently \citet{Manolopoulou2017} applied a model for determining whether a cluster rotates and, if it does, information about its rotational dynamics.

Additionally, recent efforts have been made to identify substructure within galaxy clusters.
There are many 3D, 2D, and 1D tests for substructure that have been developed in the past few decades \citep[][]{Dressler1988, West1988, West1990, Hou2009, Coziol2009}.
\citet{Pinkney1996} compared and discussed the validity of some of the earlier tests while others have applied them to SDSS clusters \citep[][]{Einasto2012}.
Recent efforts in substructure analysis have focused on identifying subpopulations based on galaxy morphological types \citep[e.g.][]{Biviano2002, Barrena2007, Chon2012, Girardi2015}.
Accounting for such substructure when measuring dynamical masses is vital in achieving accurate estimates.
For example, \citet{Old2017} has shown that almost all dynamical mass estimators overestimate cluster masses for clusters with significant dynamical substructure compared to estimates for clusters without substructure.

Furthermore, the identification and proper modeling of substructure may be important for distinguishing among competing models for the nature of dark matter.  For example, under the standard cold dark matter (CDM) paradigm, dense `cusps' form at the centers of dark matter halos \citep{Dubinski1991,Navarro1996,Navarro1997}.  In galaxy cluster halos, CDM cusps will tend to bind the brightest cluster galaxy (BCG) near the halo center.  However, recent simulations suggest that if the dark matter undergoes significant self-interactions, the subsequent unbinding of central cusps (particularly in response to major mergers) would allow BCGs to `wobble' about the cluster center \citep{Harvey2017, Kim2017}.  Such wobbles could be detected as offsets between clusters and their BCGs in the projected phase-space.  Substructure can affect the detection of such offsets, as the elements within a given substructure do not independently sample a phase space that is representative of the cluster itself.

Clearly both rotation and substructure can affect inferences about the internal dynamics of galaxy clusters.
Here we devise a framework that can account for both affects simultaneously.
This allows us to study the impacts of both phenomena on cluster mass estimates, and to marginalize over uncertainties in rotation and substructure. 
In this paper we apply this model to our own published spectroscopic observations of Abell 267 \citep[A267][]{Tucker2017}, combined with measurements from the redshift catalogue HectoSpec \citep[][]{Rines2013} to achieve a large sample.
We summarize these data sets in \S\ref{observations}.
In \S\ref{DynamicalModel} we describe the dynamical model, and we then apply the model to A267 assuming a uniform velocity dispersion (\S\ref{NoJeans}) and a dark matter halo model (\S\ref{Jeans}).
Throughout the paper we use $H_0=100\ h^{-1}\mathrm{km/s/Mpc}$ and mass density $\Omega_m=0.3$.

\section{Data}
\label{observations}

The A267 data are drawn from three separate catalogues.
The spectroscopic observations are a combination of over 1000 measured redshifts by HectoSpec \citep[HeCS,][]{Rines2013} and 223 galaxies with the Michigan/Magellan Fiber System (M2FS).
For galaxies that were observed in both data sets, we used a weighted (by inverse variance of redshift) mean of the measured redshifts.
The combination of these included 1219 galaxy redshifts with a median error of $32\mathrm{km/s}$.

The observations, data reduction, and spectral fitting model for the M2FS spectroscopy is described in detail in \citet{Tucker2017}.
We fit these spectra using a population synthesis integrated light model, that estimates line-of-sight velocity, $\vlos$, along with stellar population parameters mean age, metallicity $\feh$, chemical abundance $\afe$, and internal velocity dispersion $\disp$.
A summary of these results can be found in Table 3 of \citet{Tucker2017} and the full data product, including sky-subtracted spectra with variances, best fitting model, and samples from the posterior distribution, can be found online at \url{https://doi.org/10.5281/zenodo.831784}.

The HeCS catalog is described in detail by \citet{Rines2013} and contains redshifts for over 22,000 galaxies in over 50 different clusters.
Compared to the M2FS sample, the HeCS sample for A267 is much larger and provides wider coverage.
The M2FS sample, while smaller, provides extra dimensions of information, including mean ages and metallicities.

Both spectroscopic data sets were selected via the galaxy red sequence described in \S2.1 of \citet{Tucker2017} and shown in Fig. 1 of that paper.
We applied this same selection criteria to obtain a photometric galaxy sample from the Sloan Digital Sky Survey (SDSS) of 1849 galaxies.
The galaxies contained in the spectroscopic sample are a subset of those in the photometric sample.
Fig. \ref{A267_Positions} shows the positions of all galaxies used in this analysis.
The open markers are galaxies with only photometric observations, while the filled markers are galaxies with spectroscopically measured redshifts.
Fig. \ref{A267_vPDFs} shows the redshift distribution of galaxies used in this analysis.

Because we select galaxies via the red sequence, our inferences on cluster substructure and kinematics are biased to the quiescent galaxy population.
We note that the velocity dispersion of quiescent galaxy members has been shown in the past to be smaller than the velocity dispersion of blue members \citep[see][for example]{Zhang2012}.

\begin{figure}[t]
\centering
\includegraphics[width=\columnwidth]{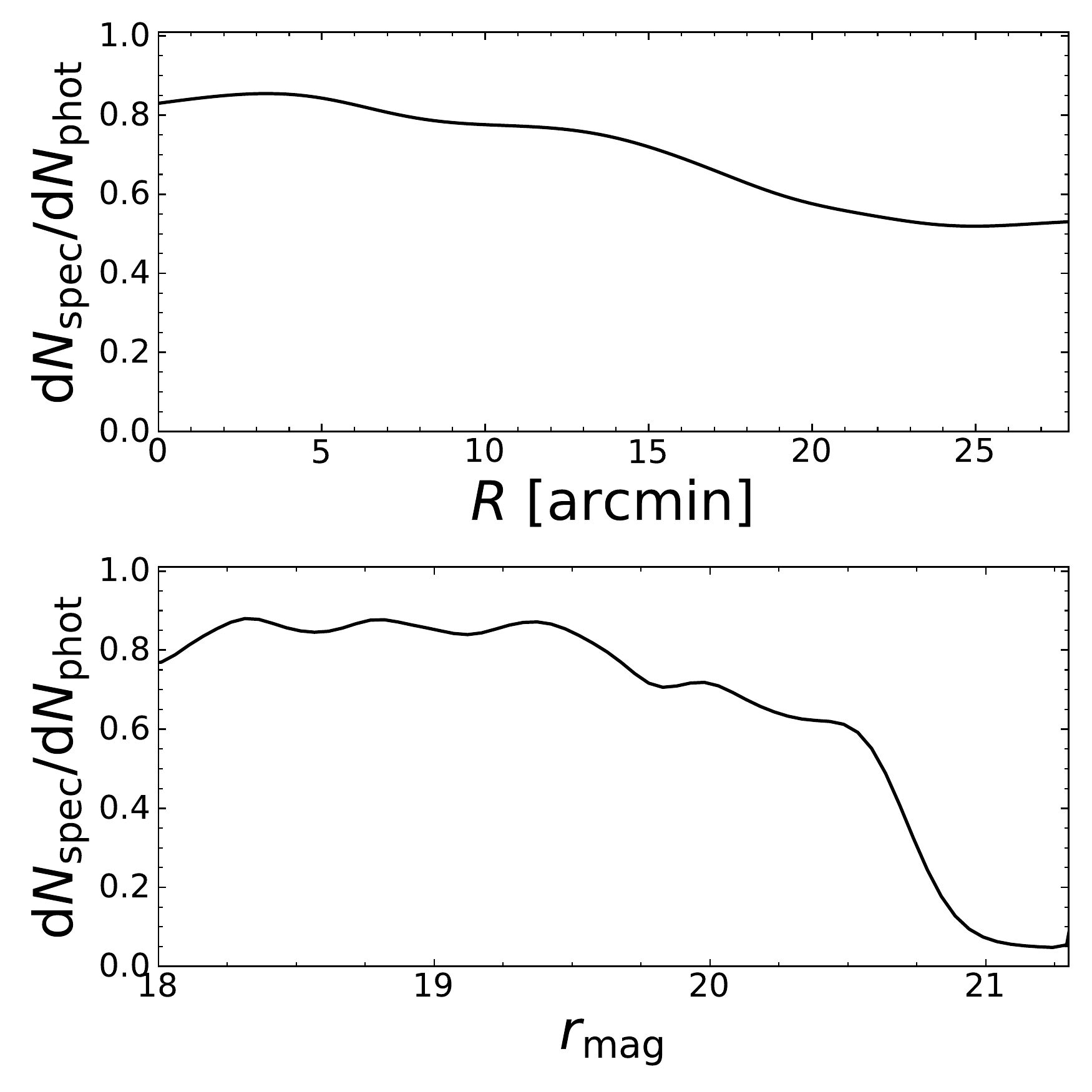}
\caption{Spectroscopic completeness as a function of radial distance (top) and r-band magnitude (bottom).}
\label{CompPlot}
\end{figure}

The spectroscopic completeness as a function of radial distance and r-band magnitude are shown in Fig. \ref{CompPlot}.
The majority of the galaxies targeted via the red sequence lie between magnitudes 18 and 21.

\section{Galaxy Cluster Mixture Model}
\label{DynamicalModel}

In this section we describe the mixture model for galaxy cluster substructure analysis.

We model the observed distribution of galaxy positions and redshifts as a random sample from several distinct galaxy populations.
We define the populations as: the main cluster population, a set of subpopulations of galaxies within the cluster, and a contamination population including both foreground and background galaxies.
We incorporate into the model the photometric observations of the galaxies from SDSS within 30' of the center of A267, along with the spectroscopic measurements of $\vlos$ from HeCS and M2FS.
We define the likelihood function that, given a set of model parameters $\bm{\theta}$, describes the observed position and velocity distribution as
\begin{equation}
\label{DynamLikelihood}
\mathcal{L}=\mathcal{L}_\mathrm{phot}\mathcal{L}_\mathrm{spec}
\end{equation}
where $\mathcal{L}_\mathrm{phot}$ is the likelihood function associated with the photometric dataset and $\mathcal{L}_\mathrm{spec}$ is associated with the spectroscopic dataset.

We model the discrete photometric sample of galaxies as being drawn independently from an underlying surface brightness profile $I(\mathbf{R})$.
Therefore, the probability of obtaining our observed photometric sample is \citep{Richardson2011}
\begin{equation}
\label{PhotLikelihood}
\mathcal{L}_\mathrm{phot}\propto\exp\left[-\int_\mathcal{R}I\left(\bm{R}\right)\bm{\mathrm{d}^2R}\right]\prod^{N_\mathrm{gal}}_iI\left(\bm{r}_i\right)
\end{equation}
where $\mathcal{R}$ is the field-of-view (FOV), $I\left(\bm{R}\right)$ is the surface brightness profile, $N_\mathrm{gal}$ is the number of galaxies observed in the photometric dataset, and $\bm{r}_i$ is the position on the sky of each galaxy.
The constant of proportionality here does not depend on the model.
For a multi population model, the surface brightness profile is the sum of the profiles for each individual population:
\begin{equation}
\label{MultiPopProfile}
I\left(\bm{R}\right)=\sum^{N_p}_{M=1}I_M\left(\bm{R}\right)
\end{equation}
where $N_p$ is the number of populations in the model.
For our multi-population model we assume the main population follows a Navarro, Frenk, and White (NFW) profile \citep{Navarro1996}: $I_1\left(\bm{R}\right)=I_\mathrm{NFW}\left(\bm{R}\right)$; contaminating galaxies follow a uniform profile: $I_2\left(\bm{R}\right)=\Sigma_\mathrm{contam}$ over the FOV.
Each of the subpopulations within the cluster follows a gaussian profile: $I_{i+2}\left(\bm{R}\right)=\Sigma_\mathrm{sub,i}\exp\left[-\frac12\left(\bm{R}-\bm{R}_i\right)^2/R_{\mathrm{sub},i}^2\right]$, where $\bm{R}_i$ is the position on the sky of the $i$-th subpoplution with scale length $R_{\mathrm{sub},i}$ and $\Sigma_\mathrm{sub,i}$ is a normalization factor.
Even though the light profile for the subpopulations may follow an NFW profile or some other non-Gaussian profile, we chose to use a Gaussian in order to increase the computational efficiency of the model because the projection of the NFW profile is not analytic.

The spectroscopic likelihood function used to describe the velocity distribution is
\begin{equation}
\label{SpecLikelihood}
\mathcal{L}_\mathrm{spec}=\prod^{N_\mathrm{spec}}_iP\left(v_i|\bm{r}_i,\bm{\theta}\right)
\end{equation}
where $N_\mathrm{spec}$ is the number of galaxies in the spectroscopic data set, $v_i$ is the velocity of each galaxy, and $P\left(v_i|\bm{R}_i,\bm{\theta}\right)$ is the probability distribution of measured line-of-sight velocity $v_i$, given position $\bm{r}_i$ and model parameters $\bm{\theta}$.
We can then marginalize this distribution over the populations and invoke Bayes' Rule to write
\begin{equation}
\label{ProbSpec1}
P\left(v_i|\bm{r}_i,\bm{\theta}\right)=\sum^{N_p}_{M=1} P\left(v_i|M,\bm{r}_i,\bm{\theta}\right) \frac{P\left(M|\bm{\theta}\right) P\left(\bm{r}_i|M,\bm{\theta}\right)}{P\left(\bm{r}_i|\bm{\theta}\right)}.
\end{equation}
The first term in the numerator is simply the number fraction of galaxies within that population: $P\left(M|\bm{\theta}\right)=f_M=N_M/N_\mathrm{tot}$.
The second term in the numerator, $P\left(\bm{r}_i|M,\bm{\theta}\right)$, is the probability for a galaxy at position $\bm{r}_i$ given the population $M$ and the model $\bm{\theta}$, which is directly proportional to the surface brightness profile of the population: $P\left(\bm{r}_i|M,\bm{\theta}\right)=2\pi r_i I_M\left(\bm{r}_i\right)/N_M$.
The denominator we can again marginalize over the populations so that $P\left(\bm{r}_i|\bm{\theta}\right)=\sum^{N_p}_{Q} P\left(\bm{r}_i|Q,\bm{\theta}\right) P\left(Q|\bm{\theta}\right)$.
And so we can re write Eq. \ref{ProbSpec1} as 
\begin{equation}
\label{ProbSpec2}
P\left(v_i|\bm{r}_i,\bm{\theta}\right)=\frac{\sum^{N_p}_{M} P\left(v_i|M,\bm{r}_i,\bm{\theta}\right) I_M\left(\bm{r}_i\right)}{\sum^{N_p}_{Q} I_Q\left(\bm{r}_i\right)}.
\end{equation}

The final probability distribution in Eq. \ref{ProbSpec2} describes the velocity distribution for a given population $M$ and position $\bm{r}_i$.
We allow for a different velocity distribution for each of the different populations incorporated into the model.
For the main cluster population, we assume that the velocity distribution follows a Gaussian profile with a rotational velocity term such that:
\begin{equation}
\label{MainVel}
P\left(v_i|M=\mathrm{main},\bm{r}_i,\bm{\theta}\right)=\frac{\exp{\left[-\frac12\frac{\left(v_i-\langle V\rangle_{267}-V_\mathrm{rot}\left(\bm{r}_i\right)\right)}{\left(\delta_i^2+\sigma(\bm{R}_i)_\mathrm{main}^2\right)}\right]}}{\sqrt{2\pi\left(\delta_i^2+\sigma(\bm{r}_i)_\mathrm{main}^2\right)}}
\end{equation}
where $\delta_i$ is the measurement uncertainty in $v_i$, $\sigma(\bm{r}_i)_\mathrm{main}$ is the velocity dispersion of the cluster at the position of each galaxy, $\langle V\rangle_{267}$ is the average velocity of the cluster, and $V_\mathrm{rot}\left(\bm{r}_i\right)$ is the rotational velocity of the cluster and is given by:
\begin{equation}
\label{MainVRot}
V_\mathrm{rot}\left(\bm{r}_i\right)=2v_\mathrm{rot}\frac{r_i}{r_i+R_\mathrm{main}}\sin\left(\theta_i-\theta_\mathrm{rot}\right)
\end{equation}
where $v_\mathrm{rot}$ is the amplitude of the rotational velocity at the NFW scale radius $R_\mathrm{main}$, and $\theta_\mathrm{rot}$ is the angle of the axis of rotation.

For each subpopulation within the cluster we assume a Gaussian velocity distribution:
\begin{equation}
\label{SubVel}
P\left(v_i|M=\mathrm{sub},\bm{r}_i,\bm{\theta}\right)=\frac{\exp{\left[-\frac12\frac{\left(v_i-\langle V\rangle_\mathrm{sub}\right)}{\left(\delta_i^2+\sigma_\mathrm{sub}^2\right)}\right]}}{\sqrt{2\pi\left(\delta_i^2+\sigma_\mathrm{sub}^2\right)}}
\end{equation}
where $\langle V\rangle_\mathrm{sub}$ and $\sigma_\mathrm{sub}$ are the mean velocity and velocity dispersion of the subpopulation.

For the contamination population we implement a weighted Gaussian smoothing kernel:
\begin{equation}
\label{ContamVel}
P\left(v_i|M=\mathrm{contam},\bm{r}_i,\bm{\theta}\right)=\frac{\sum^{N_\mathrm{spec}}_jw_j\exp\left[-\frac12\frac{\left(v_j-v_i\right)^2}{\sigma^2_\mathrm{contam}}\right]}{\sqrt{2\pi\sigma^2_\mathrm{contam}}\sum^{N_\mathrm{spec}}_jw_j}
\end{equation}
where $\scon=1000\mathrm{km/s}$ is the smoothing parameter that we set ahead of time to smooth the contamination population.
This choice for $\scon$ is arbitrary; however we confirm that our subsequent results are insensitive to the value of  $\scon$.
Furthermore, we chose $\scon=1000\ \mathrm{km/s}$ because this is the scale of galaxy cluster velocity dispersions, and foreground and background clusters dominate the contaminations population.

In order to set the weights $w_j$, we implemented an Expectation-Maximization (EM) algorithm to quickly determine a prior probability of cluster membership, $\mathcal{P}_\mathrm{mem}$, and we then set $w_j=1-\mathcal{P}_{\mathrm{mem},j}$.
Our implementation of the EM algorithm is similar to the method described by \citet{Walker2009}, with a few modifications.
First, we use the velocity distribution for the contamination population described in Eq. \ref{ContamVel} instead of the Gaussian profile described in \citet{Walker2009}; because of this, we do not update $\scon$ during the maximization step, instead we use a fixed value for $\scon$ throughout the entire algorithm.
Second, for initialization of the algorithm, we specify the seed values for $\langle V\rangle_{267}$ and $\sigma_\mathrm{main}$, the mean velocity and velocity dispersion of the cluster, respectively.
With these changes, the EM algorithm only has two parameters to update during the maximization step: $\langle V\rangle_{267}$ and $\sigma_\mathrm{main}$.

In order to test the effectiveness of this contamination component of the model, we compared the general model with a model that identifies cluster members beforehand.
We used the caustic technique \citep[]{Diaferio1999, Gifford2013} to select cluster members as the galaxies that lie within the caustic surfaces in phase space.
Then only using these galaxies we re-ran our model with a small modification; because this method already removes contaminats, we no longer include this population (Eq. \ref{ContamVel}) in the mixture model for the spectroscopic likelihood function (Eq. \ref{ProbSpec2}).
Essentially Eq. \ref{ProbSpec2} only sums over the main cluster halo and the subpopulations but no longer includes the contamination population.
The results from fitting the model in this way produce consistent results with those outlined in \S \ref{FullNoJeans} below.

For every model $\bm{\theta}$, we can evaluate the probability that each galaxy is a member of the various populations.
Given a galaxies velocity $v_i$ and position $\bm{R}_i$, the probability that it is a member of population $M$ is
\begin{multline}
\label{ProbMem}
\mathcal{P}_M=P\left(M|v_i,\bm{r}_i,\bm{\theta}\right)=\frac{P\left(M|\bm{\theta}\right)P\left(v_i,\bm{r}_i|M,\bm{\theta}\right)}{P\left(v_i,\bm{r}_i|\bm{\theta}\right)}\\
=\frac{P\left(v_i|\bm{r}_i,M,\bm{\theta}\right)I_M\left(\bm{r}_i\right)}{\sum_QP\left(v_i|\bm{r}_i,Q,\bm{\theta}\right)I_Q\left(\bm{r}_i\right)}.
\end{multline}
In the following sections we will use "probability of membership to the cluster'' to refer to the probability that an individual galaxy belongs to either the main population or any subpopulation, and we define this membership probability as $\mathcal{P}_\mathrm{mem}=1-\mathcal{P}_{M=\mathrm{contam}}$.

In order to fit this model, we use the nested sampling algorithm MultiNest \citep[][]{Feroz2009}, which simultaneously calculates the Bayesian evidence, used for model selection, and generates random samples from the posterior probability distribution.

\section{Uniform Velocity Dispersion Profile}
\label{NoJeans}

\begin{table*}
\centering
\caption{Free parameters and priors for uniform velocity dispersion model of Abell 267. For the first subpopulation, the prior range for $z_\mathrm{sub,1}$ is the same as for the main cluster: uniform between 0.22 and 0.245}
\begin{tabular}{ l l l }
\hline
\hline
Parameter &
Prior &
Description\\
\hline
$R_\mathrm{main}/R_\mathrm{max}$ & Uniform between $0$ and $1$ & Scale radius of main cluster population\\
$z_\mathrm{267}$ & Uniform between $0.22$ and $0.245$ & Redshift of A267\\
$\log_{10}\left(\sigma_\mathrm{main}/\mathrm{km\ s^{-1}}\right)$ & Uniform between $0$ and $3.5$ & Velocity dispersion of main cluster population\\
$\log_{10}\left(v_\mathrm{rot}/\mathrm{km\ s^{-1}}\right)$ & Uniform between $0$ and $3.5$ & Central rotational velocity of main cluster population\\
$\theta_\mathrm{rot}$ & Uniform between $-\pi$ and $+\pi$ & Direction of main cluster population rotational velocity\\
$\log\left[\Sigma_\mathrm{contam}/\mathrm{arcmin}^{-2}\right]$ & Uniform between $-2$ and $15$ & Surface brightness density for the uniform contamination profile\\ 
$f_\mathrm{mem}$ & Uniform between $0$ and $1$ & Number fraction of galaxies in A267 (main + subpopulations)\\ 
\hline
$R_{\mathrm{sub},i}/R_\mathrm{max}$ & Uniform between $0$ and $1$ & Scale radius of $i$-th cluster subpopulation\\
$r_\mathrm{c,sub,i}/R_\mathrm{max}$ & Uniform between $0$ and $1$ & Radial location of center of $i$-th cluster subpopulation\\
$\theta_\mathrm{c,sub,i}$ & Uniform between $0$ and $2\pi$ & Angular location of center of $i$-th cluster subpopulation\\
$z_\mathrm{sub,i}$ & Uniform between $z_\mathrm{sub,i-1}$ and $0.245$ & Redshift of $i$-th cluster subpopulation\\
$\sigma_\mathrm{sub,i}/\sigma_\mathrm{main}$ & Uniform between $0$ and $0.5$ & Velocity dispersion of $i$-th cluster subpopulation\\
\hline
$f_\mathrm{1}...f_i$ & Uniform between $0$ and $1$ & Number fraction of of galaxies in each subpopulation\\ 
\hline
\end{tabular}
\label{DynamParams}
\end{table*}

\begin{figure*}
\centering
\includegraphics[width=.45\textwidth]{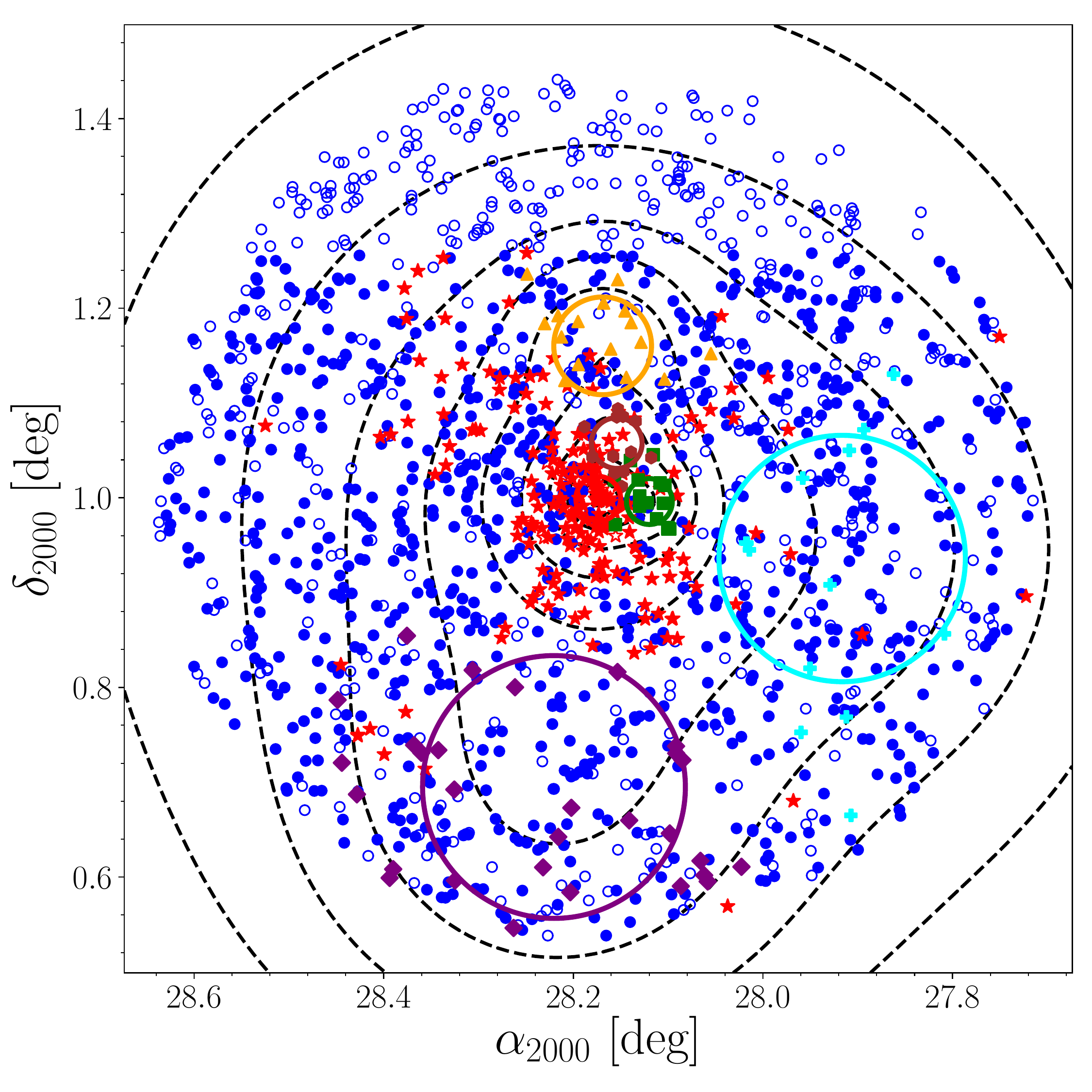}
\includegraphics[width=.45\textwidth]{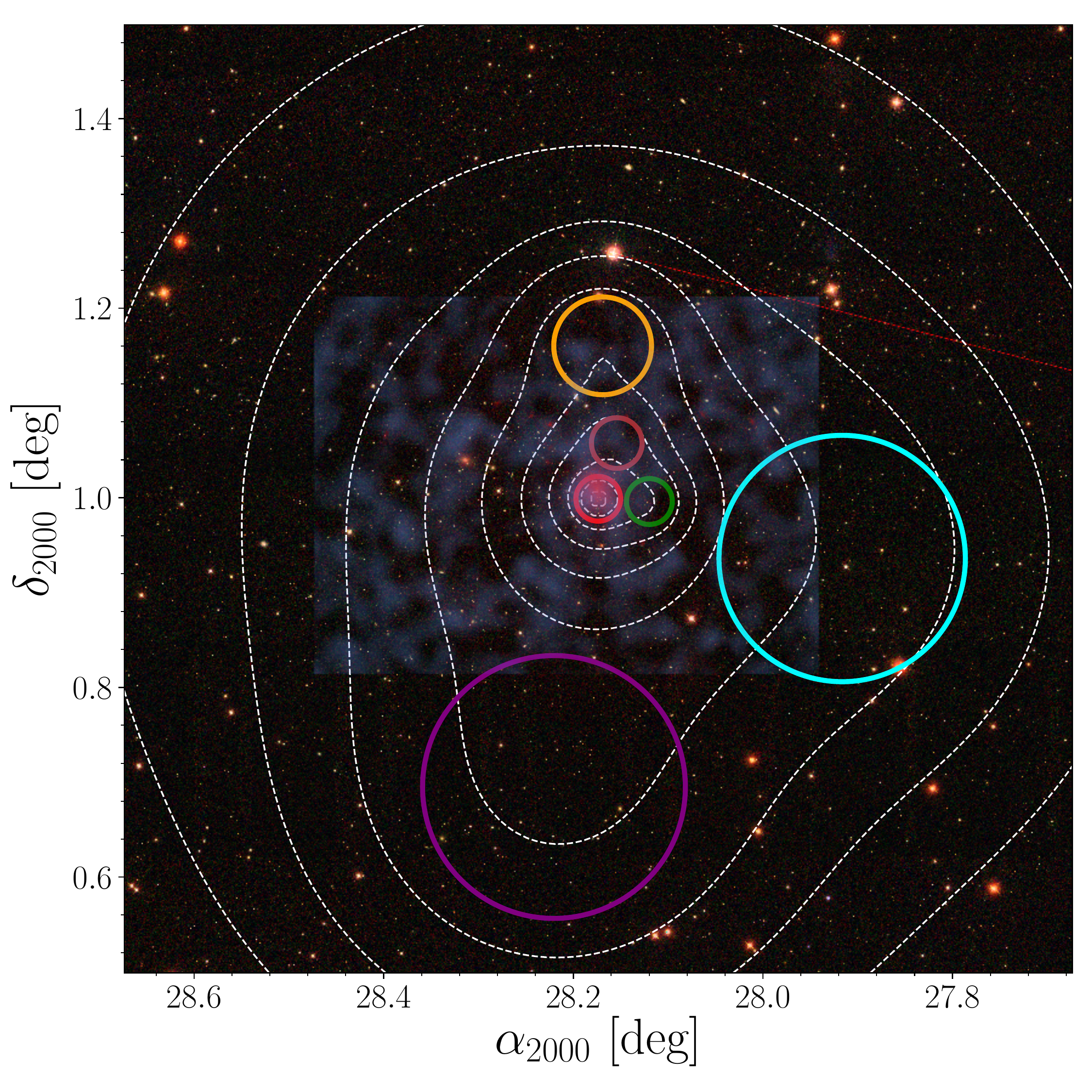}\\
\includegraphics[width=.45\textwidth]{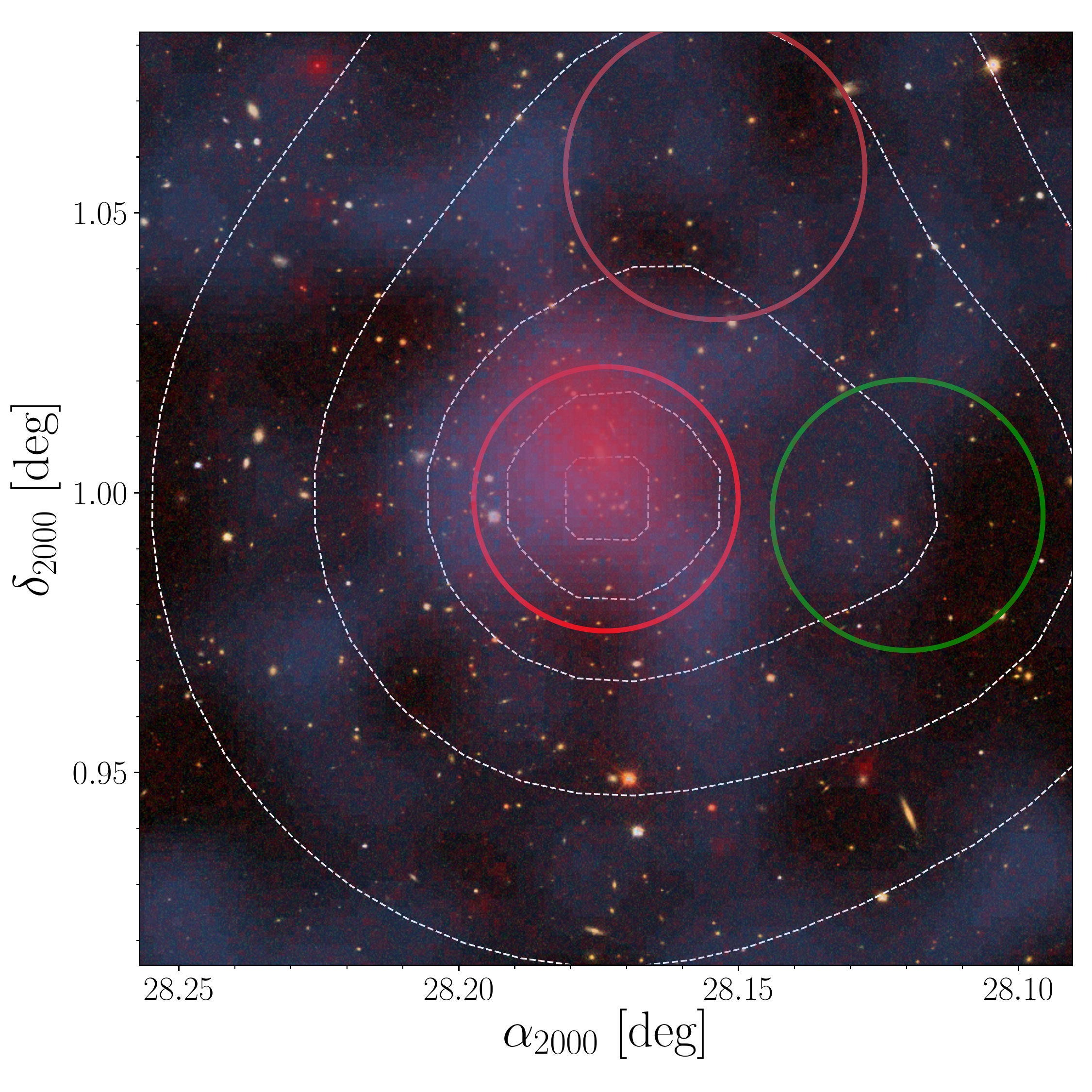}
\caption{Top left shows the positions of galaxies on the sky.  Each galaxy is colored and shaped based on which population the galaxy has the highest probability of membership: red stars are the main cluster population, green squares, purple diamonds, orange triangles, brown hexagons, and cyan crosses are for the five subpopulations, and blue circles are either foreground or background contamination galaxies. The solid red circle shows the scale radius of the main cluster population $R_\mathrm{main}$ centered on A267. The other colored circles show the scale radius $R_{\mathrm{sub},i}$ of their respective subpopulations centered on the measured center of the population. The dashed black curves show contours of equal density from the highest likelihood number density profile to the data ($\Sigma_\mathrm{main}+\Sigma_\mathrm{sub1}+\Sigma_\mathrm{sub2}+\Sigma_\mathrm{sub3}+\Sigma_\mathrm{sub4}+\Sigma_\mathrm{sub5}$). In the other two panels, we overplot these contours as well as the scale radii of the populations on top of the SDSS image center on A267, the x-ray luminosity (shown as a pink hue), and the weak-lensing signal \citep[shown in light blue,][]{Okabe2010}. The bottom panel is a zoom-in on the center of A267.}
\label{A267_Positions}
\end{figure*}

\begin{figure}
\centering
\includegraphics[width=\columnwidth]{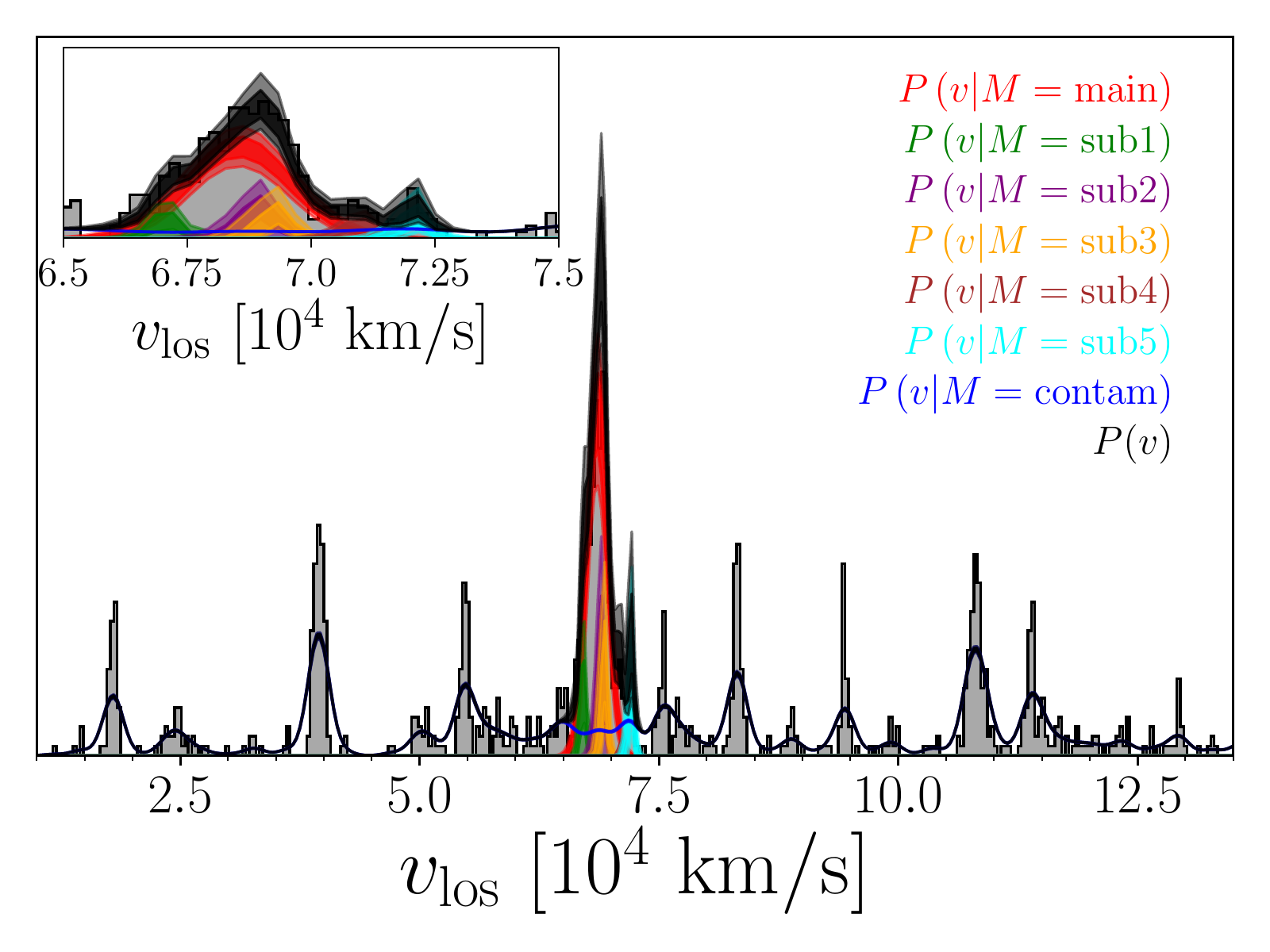}
\caption{The velocity distribution profile. The grey histograms show the profile of the galaxy redshift sample (HeCS \textit{plus} M2FS).  The red curve is the profile for the main cluster population, the green, purple, orange, brown, and cyan are for the five subpopulations, and the blue is for the contamination population.  The black curve is the sum of all of these profiles. The insert in the upper left corner shows the distribution zoomed-in on the region of redshift space around A267.}
\label{A267_vPDFs}
\end{figure}

In this section, we use a simple kinematic model in order to explore how inferences on the kinematics of A267 depend on the number of subpopulations allowed.
We do this by running six separate model fits, each model allowing an additional subpopulation (from zero to five).
The free parameters and their prior ranges used in these models are given in Table \ref{DynamParams}.

\subsection{Number of Subpopulations in A267}
\label{NSubPops}

\begin{figure}
\centering
\includegraphics[width=\columnwidth]{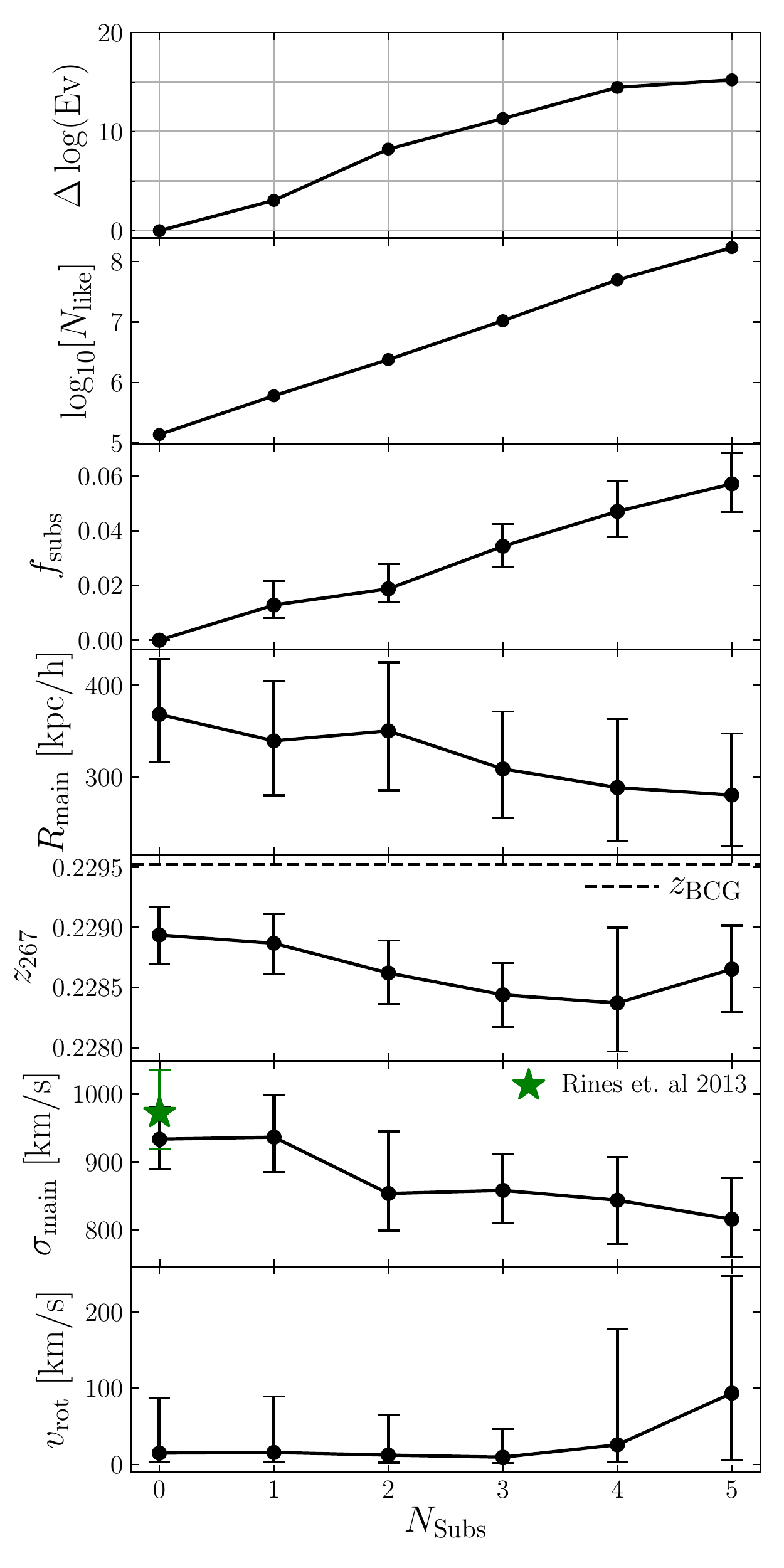}
\caption{Summary plot of the subpopulation analysis. The first panel from the top shows the evolution of the change in Bayesian evidence relative to the $N_\mathrm{Subs}=0$ model, which is commonly known as the Bayes factor. The second panel shows the number of likelihood evaluations required to adequately sample the posterior PDF of each model. Next is the number fraction of galaxies within all subpopulations. The bottom four panels are the model parameters used to describe the main cluster populations: NFW scale radius $R_\mathrm{main}$, mean cluster redshift $z_{267}$, velocity dispersion $\sigma_\mathrm{main}$, and rotational velocity $v_\mathrm{rot}$. The dashed line in the 5th panel shows the measured redshift of the BCG for A267. In the second from the bottom panel, the green star shows the velocity dispersion of A267 as measured by \citet{Rines2013}.}
\label{SubpopSummary}
\end{figure}

As discussed by \citet{Old2017}, the presence of substructure can have a significant effect on dynamical mass measurements of galaxy clusters.
In order to understand this effect on mass estimates, as well as detections of cluster rotation, we first assume a simple uniform velocity dispersion profile to explore how substructure influences these measurements.
We fit a set of six models, with each model allowing an additional subpopulation within the cluster environment (the largest number of subpopulations we fit are $N_\mathrm{Subs}=5$).
Each additional subpopulation requires 6 new free parameters to be included (central coordinates, mean velocity and velocity dispersion, scale radius, and member fraction); a summary of these parameters and the chosen prior ranges is given in Table \ref{DynamParams}.
Because all subpopulations have the same functional form, we specify some requirement so that the subpopulations are consistently ordered.
Specifically, we require each additional subpopulation to have a mean velocity larger than that of the previous subpopulation, i.e. $\langle V\rangle_\mathrm{sub1}<\langle V\rangle_\mathrm{sub2}...$ etc.
This requirement is imposed on the prior ranges and is reflected in the row for $z_\mathrm{sub,i}$ in Table \ref{DynamParams}.

Fig. \ref{SubpopSummary} shows a summary of the main results from this analysis.
In the top panel we show the evolution of the change in the log evidence for each model relative to the $N_\mathrm{Subs}=0$ model.
This value is frequently referred to as the Bayes factor, and it is commonly used for model selection.
The larger the Bayes factor, the more significant the evidence is that new model is ``better" than the previous model, accounting for differences in model complexity.
According to \citet{Kass1995}, a Bayes factor ($\Delta\log(\mathrm{Ev})$) between 3 and 5 indicates ``strong'' evidence and if this factor exceeds 5, then the new model is very strongly favored. 
The Bayes factor is consistently $>3$ for the $N_\mathrm{Subs}=$ 1, 2, and 3 models, which indicates that each of these models is strongly favored over the model with one less subpopulation ($N_\mathrm{Subs}=$ 0, 1, and 2 respectively).
However, the $N_\mathrm{Subs}=$ 4 and 5 (with Bayes factors $<3$) are ``slightly positive" or ``not worth more than a bare mention" compared to the model with one less subpopulation.
Thus, the Bayes factor appears to favor the $N_\mathrm{Subs}=$ 3 or 4 model.

The second panel in Fig. \ref{SubpopSummary} shows the number of likelihood evaluations needed to adequately sample the posterior PDF of each model.
As expected for models with increasing number of free parameters, the required number of likelihood evaluations increases exponentially.

The third panel in Fig. \ref{SubpopSummary} shows the number fraction of galaxies in all subpopulations.
This panel gives an idea of how many galaxies are added to the subpopulations with increasing number of subpopulations.

\begin{table*}
\centering
\caption{Mean values and standard deviations of 1D posterior PDFs for A267 free parameters in the uniform velocity dispersion model}
\begin{tabular}{ c c c c c c c c c }
\hline
\hline
 & $R_0/\mathrm{Mpc\ h^{-1}}$ & $\alpha_{2000}/\mathrm{deg}$ & $\delta_{2000}/\mathrm{deg}$ & $z$ & $\sigma/\mathrm{km\ s^{-1}}$ & $v_\mathrm{rot}/\mathrm{km\ s^{-1}}$ & $\theta_\mathrm{rot}/\mathrm{rad}$ & $f_\mathrm{mem}$\\
\hline
$\mathrm{Main}$ & $0.29\pm0.06$ & $28.174$ & $0.999$ & $0.2287\pm0.0004$ & $818\pm59$ & $121\pm114$ & $1.21\pm1.63$ & $0.157\pm0.014$\\
$\mathrm{Sub1}$ & $0.45\pm0.63$ & $28.114\pm0.040$ & $1.003\pm0.003$ & $0.2238\pm0.0010$ & $195\pm72$ & & & $0.009\pm0.004$\\
$\mathrm{Sub2}$ & $1.54\pm0.46$ & $28.205\pm0.006$ & $0.712\pm0.058$ & $0.2298\pm0.0009$ & $269\pm79$ & & & $0.018\pm0.007$\\
$\mathrm{Sub3}$ & $0.71\pm0.30$ & $28.160\pm0.004$ & $1.168\pm0.046$ & $0.2311\pm0.0005$ & $225\pm77$ & & & $0.012\pm0.005$\\
$\mathrm{Sub4}$ & $0.33\pm0.08$ & $28.157\pm0.003$ & $1.056\pm0.011$ & $0.2367\pm0.0004$ & $247\pm63$ & & & $0.008\pm0.003$\\
$\mathrm{Sub5}$ & $1.52\pm0.37$ & $27.926\pm0.036$ & $0.925\pm0.011$ & $0.2403\pm0.0004$ & $189\pm72$ & & & $0.011\pm0.004$\\
\hline
\end{tabular}
\label{NoJeansResults}
\end{table*}

In the other four panels of Fig. \ref{SubpopSummary} we show the evolution of free parameters describing the main cluster: NFW scale radius $R_\mathrm{main}$, mean cluster redshift $z_{267}$, velocity dispersion $\sigma_\mathrm{main}$, and cluster rotational velocity $v_\mathrm{rot}$.
The second from the bottom panel ($\sigma_\mathrm{main}$) shows the evolution of velocity dispersion, or in other words the mass of the cluster.
For comparison, we include the velocity dispersion for A267 measured by \citet{Rines2013}, which is calculated by first identifying cluster members via the Caustic technique \citep[][]{Diaferio1997} and calculating the dispersion of the members about the mean cluster redshift (also determined via the Caustic method).
The Caustic method does not explicitly consider the effects of substructure (unless it is evident in the plane of $v_\mathrm{los}-R$), so we compare it to to our measurement assuming $N_\mathrm{Subs}=0$, finding good agreement.
As the number of subpopulations increases, the velocity dispersion decreases; furthermore, the velocity dispersion decreases by $\sim100\ \mathrm{km/s}$ from $N_\mathrm{Subs}=0$ to $N_\mathrm{Subs}=4$.
Assuming that $M_\mathrm{vir}\propto\sigma^2$ this difference could contribute a $\sim20\%$ error in dynamical mass estimates for A267.

The inflation of velocity dispersion due to the presence of substructure is not a new result.
\citet{Beers1982} studied the dynamics Abell 98, and showed that the cluster was sub-structured with two distinct components.
Furthermore, by using a two component model to fit the cluster dynamics, they showed that failure to recognize this substructure inflates the velocity dispersion and hence the mass-to-light ratio of the cluster.
\citet{Geller1984} obtain a similar result for the Cancer Cluster.
What is new here is the ability to evaluate the number of substructures and estimate cluster mass while marginalizing over uncertainty in the substructure parameters.

As the number of subpopulations increases, the scale radius $R_\mathrm{main}$ decreases.
This trend is consistent with the mass of the main cluster also decreasing.
Another trend of note is that the mean cluster redshift $z_{267}$ decreases.
The first two subpopulations identified by the model are subpopulations with redshifts larger than $z_{267}$, therefore, once these populations are accounted for, we would expect $z_{267}$ to also decrease.
Furthermore, we also include the redshift of the BCG of A267.
Clearly, the redshift of the BCG differs from the measured mean cluster redshift (this offset is on the order of $100$ km/s).
For all models considered in this analysis, we do not detect any significant cluster rotation.

We also note that our models are consistent with each other as we increase the number of subpopulations.
In other words, if a subpopulation is identified in the $N_\mathrm{subs}=1$ this same subpopulation will remain in the analysis when $N_\mathrm{subs}=5$.
This is important because each model is independent of the previous, there is no guarantee that the identification of the substructure will be consistent.

\subsection{Comparison to standard tests for substructure}

\begin{figure}
\centering
\includegraphics[width=\columnwidth]{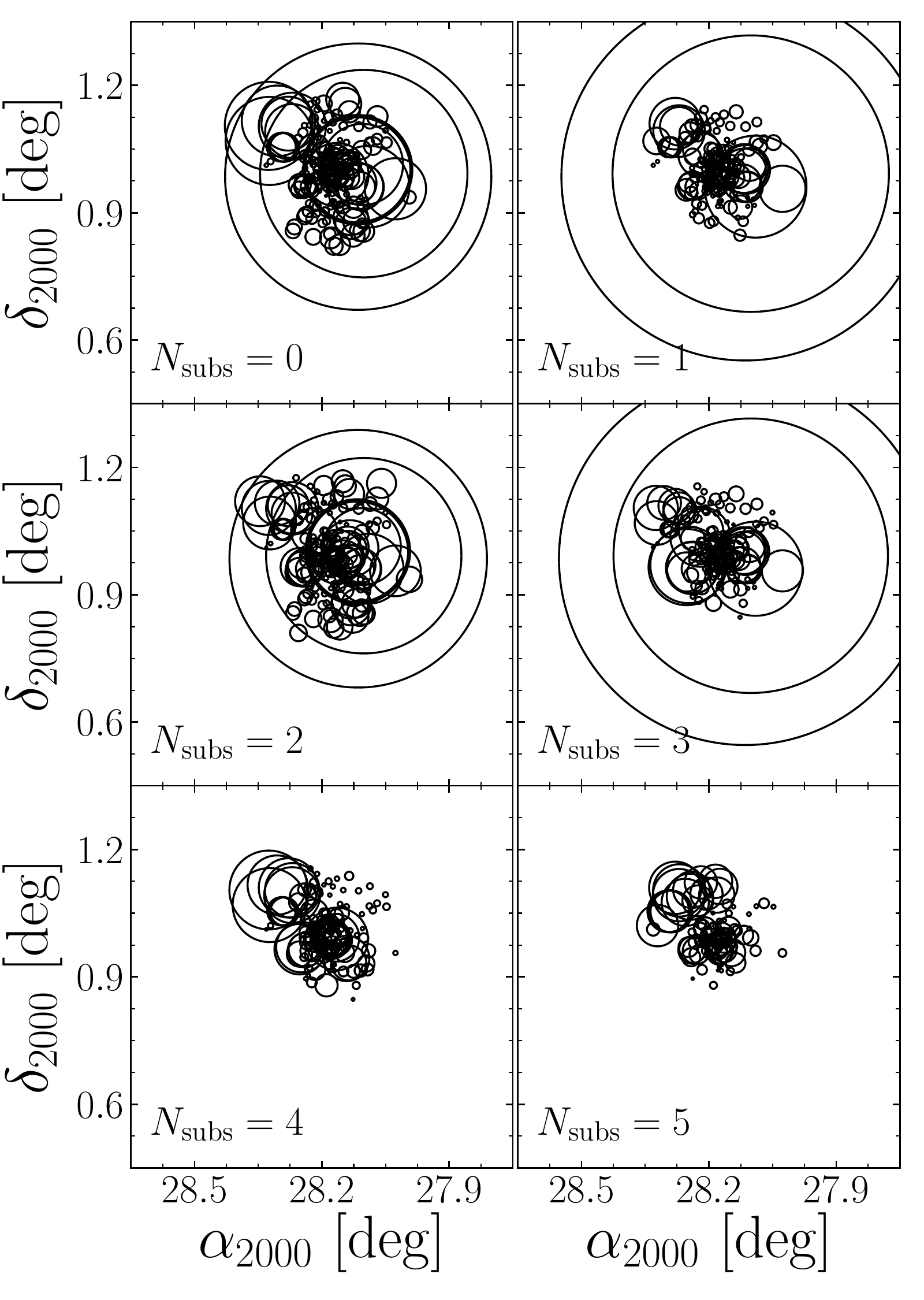}
\caption{``Bubble plot" for the $\Delta$-statistic. Each member galaxy is plotted with a circle whose size is proportional to $\delta_i$ (Eq. \ref{delta-stat}). Regions of large circles show areas with high probability of substructure. From left to right and top to bottom we increase the number of subpopulations which is given in the bottom left of each panel.}
\label{DeltaBubble}
\end{figure}

A commonly used statistical test for substructure is known as the $\Delta$-statistic and was developed by \citet{Dressler1988}.
The $\Delta$-statistic looks for deviations in the local velocity from the global velocity of the cluster.
First, for each galaxy one calculates the mean local velocity $v_\mathrm{local}$ and local dispersion $\sigma_\mathrm{local}$ of the $n$ nearest neighbors to the galaxy, where typically $n\sim\sqrt{N_\mathrm{tot}}$.
This local velocity and dispersion is compared to the global velocity $\langle V\rangle_{267}$ and dispersion $\sigma_\mathrm{main}$ of the cluster quantified by
\begin{equation}
\label{delta-stat}
\delta_i^2=\left(n+1\right)\left[\left(v_\mathrm{local}-\langle V\rangle_{267}\right)^2+\left(\sigma_\mathrm{local}-\sigma_\mathrm{main}\right)^2\right]/\sigma_\mathrm{main}^2.
\end{equation}
The full $\Delta$-statistic is the sum of $\delta_i$ over all galaxies $N_\mathrm{tot}$.

Fig. \ref{DeltaBubble} shows a ``Bubble Plot", a commonly used representation of the $\Delta$-statistic.
Each galaxy's ``bubble" is sized by that galaxy's $\delta_i$ value given by Eq. \ref{delta-stat}.
In each panel we show the progression of this plot for increasing number of subpopulations.
We first apply a hard cut on the probability of membership to the main cluster $\mathcal{P}_\mathrm{main}$ and only show galaxies with $\mathcal{P}_\mathrm{main}>0.9$.
We also only show the results for the maximum likelihood model for each number of subpopulations.
Fig. \ref{DeltaBubble} clearly shows that as more subpopulations are removed from the main cluster, the sizes of the ``bubbles'' decrease because our substructure analysis identifies the same substructure that affects the $\Delta$-statistic.

The previous two subsections give a glimpse into the effects substructure has on the analysis of cluster kinematics.
The Bayes factor appears to favor the $N_\mathrm{subs}=3$ or $4$ models, while the $\delta$-statistic requires at least $N_\mathrm{subs}=4$ in order to drastically reduce the size of the $\delta$-bubbles.

\subsection{Five subpopulation model of A267}
\label{FullNoJeans}

\begin{figure}[t]
\centering
\includegraphics[width=\columnwidth]{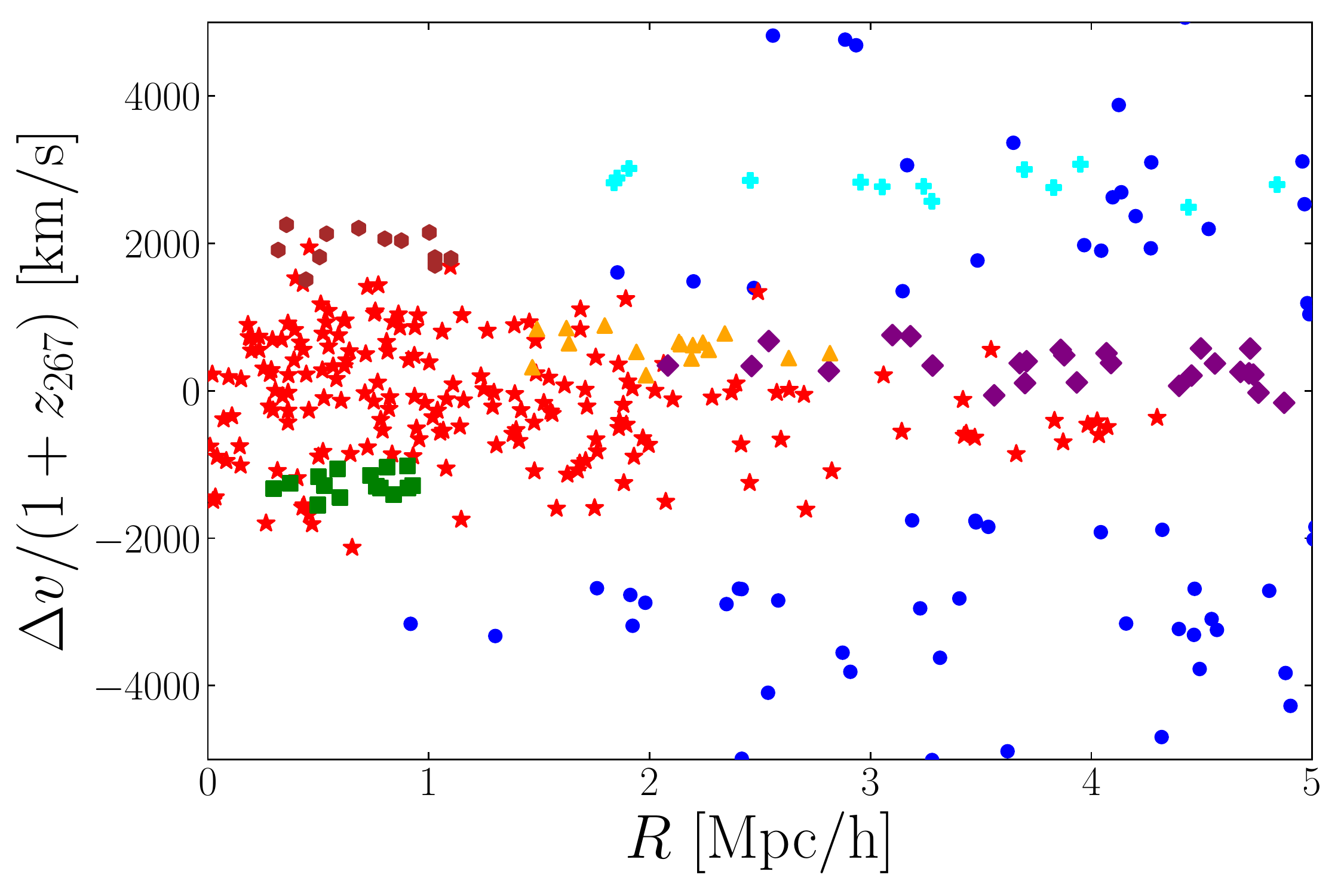}
\caption{Rest frame cluster centric velocity versus radius. Galaxies are colored and shaped in the same manor as described in Fig. \ref{A267_Positions}.}
\label{A267_PhaseSpace}
\end{figure}

In order to explore the features of the galaxy subpopulations identified in our analysis, Fig. \ref{A267_Positions} portrays the highest-likelihood model obtained for the case of $N_\mathrm{subs}=5$.
We show in the top left panel the positions of the galaxies on the sky colored by each galaxy's most likely population membership.
The red stars are galaxies that are most probable members to the main cluster population, and the green squares, purple diamonds, orange triangles, brown hexagon, and cyan crosses show the positions of the galaxies that are most probable members of the five subpopulations.
The blue circles are foreground and background contamination galaxies.
The filled symbols are galaxies with spectroscopic redshifts, while the open symbols have just photometry.
The solid red circle shows the scale radius $R_\mathrm{main}$ of the cluster, which for an NFW profile gives a size to the core of the cluster.
The other colored circles show the scale size of their respective subpopulation.
The black dashed contours show curves of equal surface number density of the cluster as a whole; in other words, these show the contours of equal density of the main cluster population plus the subpopulations.

The other two panels show an overlay image of our analysis (the colored circles and white contours) with the SDSS mosaic \citep[][]{SDSSDR14}, the weak lensing signal in blue \citep[][]{Okabe2010}, and x-ray luminosity in pink (XMM-Newton objid 0084230401).
The bottom most panel is a zoom-in on the central core of A267.

The grey histogram in Fig. \ref{A267_vPDFs} shows the velocity distribution of the galaxies from our sample.
It is easy to see the strong peak at $v_\mathrm{los}\sim69000\ \mathrm{km/s}$ ($z\sim0.23$) which is the redshift of A267.
The colored over plotted curves show the posterior of the velocity distribution profiles from the model.
The red curve shows the velocity distribution for the main cluster population (Eq. \ref{MainVel}), the green, purple, orange, brown, and cyan curves are the velocity distributions of the subpopulations (Eq. \ref{SubVel}), the blue curve shows the smoothed contamination velocity profile (Eq. \ref{ContamVel}), and the black curve shows the sum of the all populations which fits the true distribution nicely.
The insert in the top left shows a zoom-in on the region of velocity space around the mean cluster velocity to better show the distributions of each of the populations.
The histograms of the redshift catalogue show a large amount of background and foreground clusters, which our model accounts for (see Eq. \ref{ContamVel}).

Fig. \ref{A267_PhaseSpace} shows how the various sub-groups populate the projected phase space for A267.
As in Fig. \ref{A267_Positions}, the points in Fig. \ref{A267_PhaseSpace} are colored depending on the subpopulation they are most likely members of.
In combination with Fig. \ref{A267_Positions} and Fig. \ref{A267_vPDFs}, we give a simplistic view of the 3D phase space of A267.
Our substructure analysis has constrained five distinct populations.
The first population (colored in green in these figures) is a compact populations of galaxies near the cluster center and with a mean redshift $\sim1500\ \mathrm{km/s}$ smaller than the mean redshift of A267.
This subpopulation is clearly seen in Fig. \ref{DeltaBubble} because these galaxies have a large $\delta$ value.
The second subpopulation (colored in purple) is a diffuse population of galaxies to the south of the main cluster with a mean redshift very similar to the mean cluster redshift. 
The third subpopulation (colored in orange), which also has a mean redshift similar to the mean cluster redshift, is more compact yet still relatively diffuse compared to the first subpopulation and is located just north of the cluster center.
The fourth subpopulation (colored in brown) is a compact population of galaxies near the cluster core with a mean velocity of $\sim71000\ \mathrm{km/s}$, about $\sim2000\ \mathrm{km/s}$ larger than the mean cluster redshift.

The last subpopulation (colored in cyan) is a long extended population of galaxies with mean velocity of $\sim72000\ \mathrm{km/s}$ located to the west of the cluster center, and extends almost the same length on the sky as the extended cluster halo.
It is unclear whether this population is more properly considered as a loosely bound member of A267, or as a relatively nearby component of the background.
However, this distinction is unimportant for our analysis of the kinematics of A267's main component.
As long as this group is identified either as a subpopulation or background it does not contribute to the measured velocity dispersion of A267.
Table \ref{NoJeansResults} gives the mean and standard deviation for each 1D posterior for each free parameter; for some parameters, we converted these posteriors into real units.

The scale radii cited in Table \ref{NoJeansResults} have different meanings for the main and sub populations.
The scale radius of the main cluster $R_\mathrm{main}$ is the scale radius of an NFW profile, which is roughly equal to the size of the core of the cluster.
On the other hand, the scale radii for the subpopulations $R_\mathrm{sub,i}$ is the scale radius for a Gaussian profile, which corresponds to the radius that encompasses roughly $\sim68\%$ of the subpopulation's member galaxies.
Therefore, the scale radii for the subpopulations is inherently larger even though these populations are actually smaller than the main cluster population.

In this modeling, we have not attempted to incorporate stellar-population parameters estimated in \citet{Tucker2017} (mean galaxy age, metallicity, and alpha-enrichment), because we only have these parameter estimates for a small sub-sample of the galaxies included in this analysis.
However, including these parameters within the multi-population mixture model would potentially give more power to separate subpopulations, and would enable more detailed studies of galaxy evolution within the cluster environment.

\section{Dark Matter Halo Model}
\label{Jeans}

Thus far we have assumed that A267's velocity distribution is independent of radius.
In the following subsections, we describe our procedure and results for fitting a dark matter halo model and corresponding velocity dispersion profile to the main cluster population of A267.
We will first describe the theoretical framework for calculating the velocity dispersion profile as a proxy for cluster mass using the spherical Jeans equation, and then how we implement this technique for A267.

\subsection{Jeans Analysis}
In order to measure cluster mass, we assume that the galaxies within the main cluster population sample a single, pressure-supported halo that is dynamically relaxed and traces an underlying dark matter dominated gravitational potential.
With the additional assumption of spherical symmetry, the mass profile, $M(r)$, of the dark matter halo relates to the galaxy distribution function via the Jeans equation:
\begin{equation}
\label{Jeans Equation}
\frac1\nu\frac{d}{dr}\left(\nu\sigma_r^2\right)+2\frac{\beta\sigma_r^2}{r}=-\frac{GM(r)}{r^2}
\end{equation}
where $\nu(r)$ is the three-dimensional galaxy number density, $\sigma_r^2(r)$ is the radial velocity dispersion, and $\beta\equiv1-\sigma_\theta^2/\sigma_r^2$ is the orbital anisotropy.
Using cosmological dark matter only simulations, \citet{Wojtak2013} showed that the velocity anisotropy for cluster-sized halos ($10^{14}-10^{15}h^{-1}M_\odot$) is roughly constant with radius at a value $\beta\sim0.4$.
According to \citet{Binney2008}, for the special case of constant, non-zero anisotropy, the Jeans equation has the simple solution:
\begin{equation}
\label{Jeans Solution}
\nu\sigma_r^2=Gr^{-2\beta}\int_r^\infty s^{2\beta-2}\nu(s)M(s)ds.
\end{equation}
And by projecting along the line of sight, we can relate the mass profile to the observable profiles of the projected galaxy number density $I(R)$ and velocity dispersion profile $\sigma_p(R)$ by
\begin{equation}
\label{Jeans Projected}
\sigma_p^2(R)=\frac{2}{I(R)}\int_R^\infty\left(1-\beta\frac{R^2}{r^2}\right)\frac{\nu\sigma_r^2r}{\sqrt{r^2-R^2}}dr.
\end{equation}
And so, by plugging in Eq. \ref{Jeans Solution} into Eq. \ref{Jeans Projected}, specifying an underlying dark matter halo model $M(R)$, and adopting a profile for $I(R)$, we can determine the velocity dispersion and mass profiles of the cluster.

For the galaxy number density, we assume that galaxies trace out an underlying NFW profile such that $\nu(r)=\nu_0R_\mathrm{main}^3r^{-1}(R_\mathrm{main}+r)^{-2}$.
We then project this to obtain the surface brightness profile \citep{Binney2008}:
\begin{equation}
\label{projection}
I(R)=2\int_R^\infty\frac{\nu(r)rdr}{\sqrt{r^2-R^2}}.
\end{equation}
And for the dark matter halo, we adopt another NFW profile $\rho(r)$ such that the mass profile is given by $M(R)=4\pi\int_0^Rr^2\rho(r)dr$.
We would like to note that Eq. \ref{projection} is not analytic for an NFW profile for $R<R_\mathrm{main}$, and so this integral must be calculated numerically.

For simple anisotropy profiles, we can rewrite the combination of Eq. \ref{Jeans Solution} and \ref{Jeans Projected} as \citep{Mamon2005}
\begin{equation}
\label{Jeans Simple}
\sigma_p^2(R)=\frac{2G}{I(R)}\int_R^\infty K\left(\frac{r}{R},\frac{r_a}{R}\right)\nu(r)M(r)\frac{dr}{r}
\end{equation}
where the kernel K depends on the choice of anisotropy, and are given for five anisotropy models in appendix 2 of \citet{Mamon2005}.
For A267, we used a constant anisotropy model.
Although it is not the most physically motivated model, we use a Gaussian velocity profile \citep[similar to][]{Mamon2013} because it is easy to implement numerically and is a fairly good approximation for the observed profile of galaxy clusters.

\subsection{Results for A267}

\begin{table*}
\centering
\caption{Free parameters and priors for dynamical halo model of Abell 267}
\begin{tabular}{ l l l }
\hline
\hline
Parameter &
Prior &
Description\\
\hline
$R_\mathrm{main}/R_\mathrm{max}$ & Uniform between $0$ and $1$ & Scale radius of main cluster population\\
$z_\mathrm{267}$ & Uniform between $0.22$ and $0.245$ & Redshift of A267\\
$\log_{10}(M_{200}/\mathrm{Mpc\ h^-1})$ & Uniform between 13 and 16 & Virial mass of dark matter halo\\
$\log_{10}(c_{200})$ & Uniform between 0 and 2 & Concentration of dark matter halo $c_{200}=r_{200}/r_\mathrm{DM}$\\
$-\log_{10}(1-\beta)$ & Uniform between -1 and 1 & Constant anisotropy $\beta$ of velocity dispersion profile \\
$\log\left[\Sigma_\mathrm{contam}/\mathrm{arcmin}^{-2}\right]$ & Uniform between $-2$ and $15$ & Surface brightness density for the uniform contamination profile\\ 
$f_\mathrm{mem}$ & Uniform between $0$ and $1$ & Number fraction of galaxies in A267 (main + subpopulations)\\ 
\hline
$R_{\mathrm{sub},i}/R_\mathrm{max}$ & Uniform between $0$ and $1$ & Scale radius of $i$-th cluster subpopulation\\
$r_\mathrm{c,sub,i}/R_\mathrm{max}$ & Uniform between $0$ and $1$ & Radial location of center of $i$-th cluster subpopulation\\
$\theta_\mathrm{c,sub,i}$ & Uniform between $0$ and $2\pi$ & Angular location of center of $i$-th cluster subpopulation\\
$z_\mathrm{sub,i}$ & Uniform between $z_\mathrm{sub,i-1}$ and $0.245$ & Redshift of $i$-th cluster subpopulation\\
$\sigma_\mathrm{sub,i}/\mathrm{km\ s^{-1}}$ & Uniform between $0$ and $500$ & Velocity dispersion of $i$-th cluster subpopulation\\
\hline
$f_\mathrm{1}...f_i$ & Uniform between $0$ and $1$ & Number fraction of of galaxies in each subpopulation\\ 
\hline
\end{tabular}
\label{DynamParams2}
\end{table*}

\begin{table*}
\centering
\caption{Mean values and standard deviations of 1D posterior PDFs for A267 free parameters in the Dark Matter Halo Model}
\begin{tabular}{ c c c c c c c c c c }
\hline
\hline
 & $R_0/\mathrm{Mpc\ h^{-1}}$ & $\alpha_{2000}/\mathrm{deg}$ & $\delta_{2000}/\mathrm{deg}$ & $z$ & $M_{200}/10^{14}M_\odot\ h^{-1}$ & $\log_{10}(c_{200})$ & $\beta$ & $\sigma/\mathrm{km\ s^{-1}}$\\
\hline
$\mathrm{Main}$ & $0.28\pm0.06$ & $28.174$ & $0.999$ & $0.2283\pm0.0003$ & $6.77\pm1.06$ & $0.61\pm0.39$ & $-0.67\pm1.75$ & \\
$\mathrm{Sub1}$ & $1.04\pm1.34$ & $28.070\pm0.095$ & $0.984\pm0.014$ & $0.2238\pm0.0020$ & & & & $179\pm89$\\
$\mathrm{Sub2}$ & $1.47\pm0.60$ & $28.205\pm0.007$ & $0.707\pm0.070$ & $0.2302\pm0.0011$ & & & & $234\pm80$\\
$\mathrm{Sub3}$ & $0.69\pm0.47$ & $28.170\pm0.001$ & $1.164\pm0.042$ & $0.2313\pm0.0007$ & & & & $205\pm70$\\
$\mathrm{Sub4}$ & $0.32\pm0.12$ & $28.155\pm0.004$ & $1.058\pm0.012$ & $0.2368\pm0.0004$ & & & & $240\pm62$\\
$\mathrm{Sub5}$ & $1.52\pm0.35$ & $27.932\pm0.034$ & $0.903\pm0.013$ & $0.2403\pm0.0004$ & & & & $194\pm74$\\
\hline
\end{tabular}
\label{JeansResults}
\end{table*}

\begin{figure}
\centering
\includegraphics[width=\columnwidth]{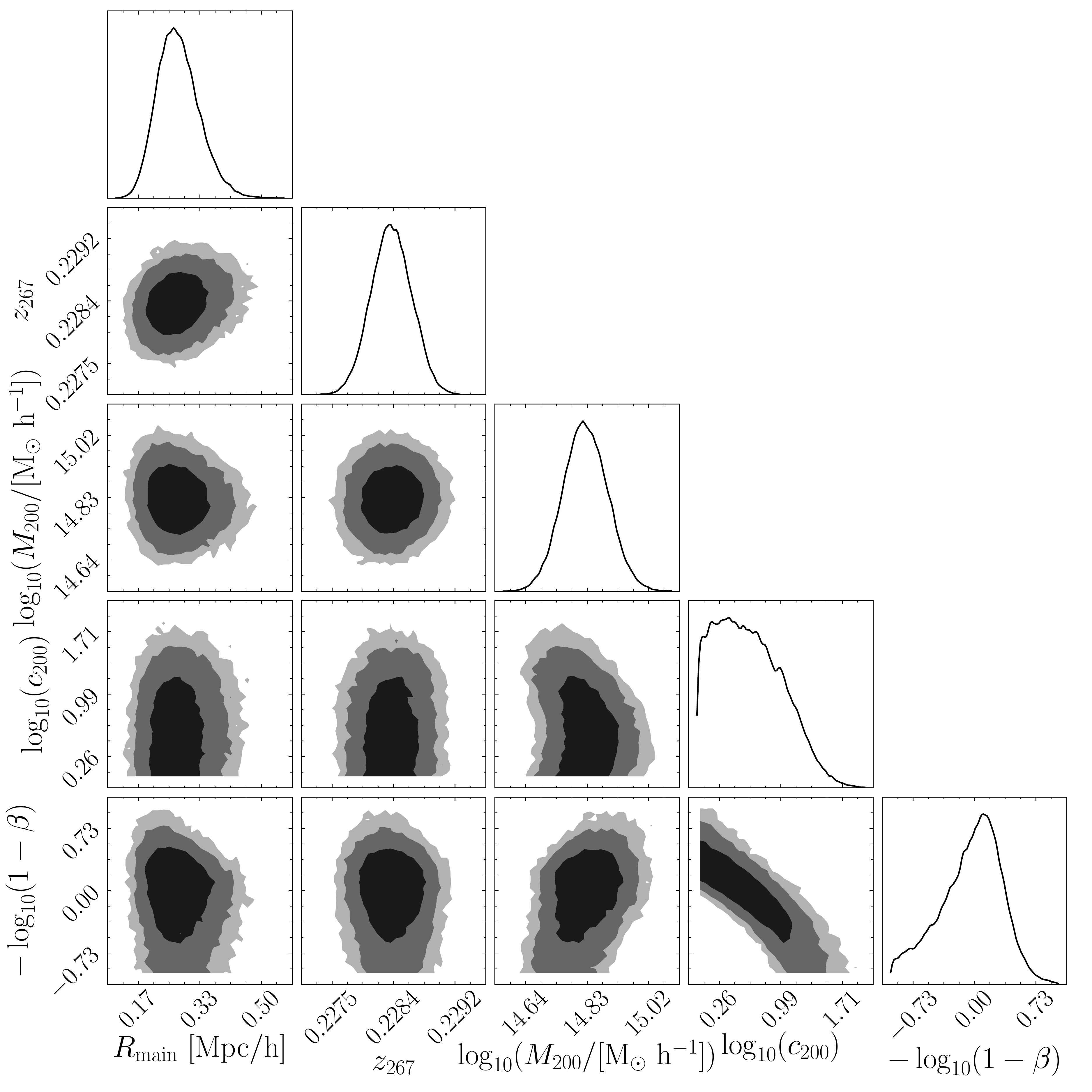}
\caption{Posterior PDFs of parameters specifying the dark matter halo of A267, using the Jeans Equation analysis described in \S\ref{Jeans}. We show the scale radius of the NFW light profile $R_\mathrm{main}$, mean cluster redshift $z_{267}$, virial mass and concentration of the dark matter halo $M_{200}$ and $c_{200}$, and velocity anisotropy $\beta$. We also show the 1, 2, and 3$\sigma$ contours for the 2D posteriors.}
\label{MainPDFs}
\end{figure}

\begin{figure}
\centering
\includegraphics[width=\columnwidth]{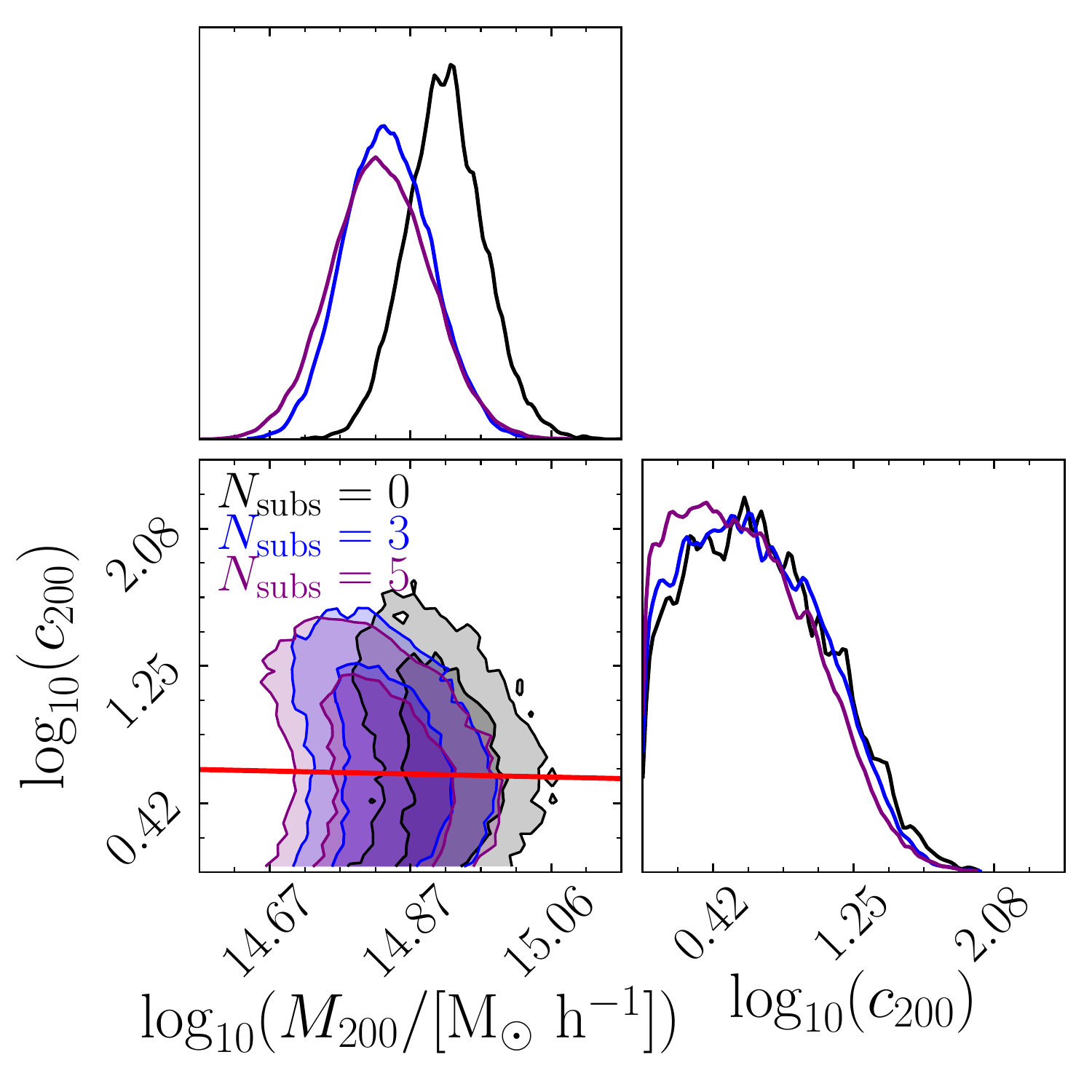}
\caption{Mass-Concentration posteriors from the dark matter halo model. Each color represents a model that allows for a different number of subpopulations. The red curve shows the $M_{200}-c_{200}$ relation from \citet{Dutton2014}.}
\label{MassConcentration}
\end{figure}

\begin{figure}
\centering
\includegraphics[width=\columnwidth]{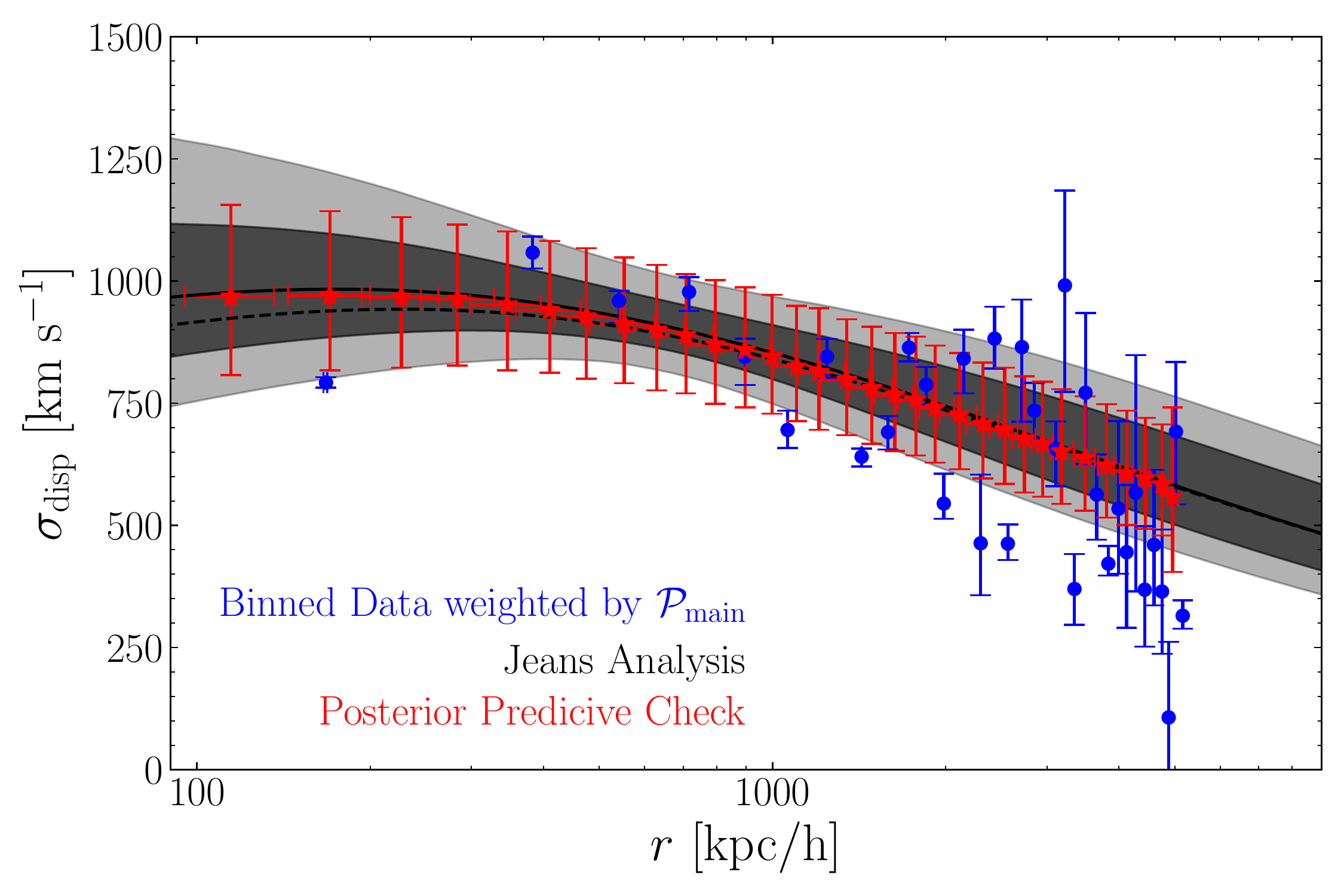}
\caption{Line-of-sight velocity dispersion profile for A267.The dark and lighter regions correspond to $1\sigma$ and $2\sigma$ of the posterior, respectively. The solid black line is the median posterior curve, while the dotted black line corresponds to $\sigma_\mathrm{disp}(r)$ for the highest likelihood model. The blue points show the velocity dispersion of binned galaxies weighted by each galaxies probably of membership to the main cluster population $\mathcal{P}_\mathrm{main}$. The red points show the binned velocity dispersion calculated from simulated data at each point in the posterior.}
\label{DispProf}
\end{figure}

\begin{figure}
\centering
\includegraphics[width=\columnwidth]{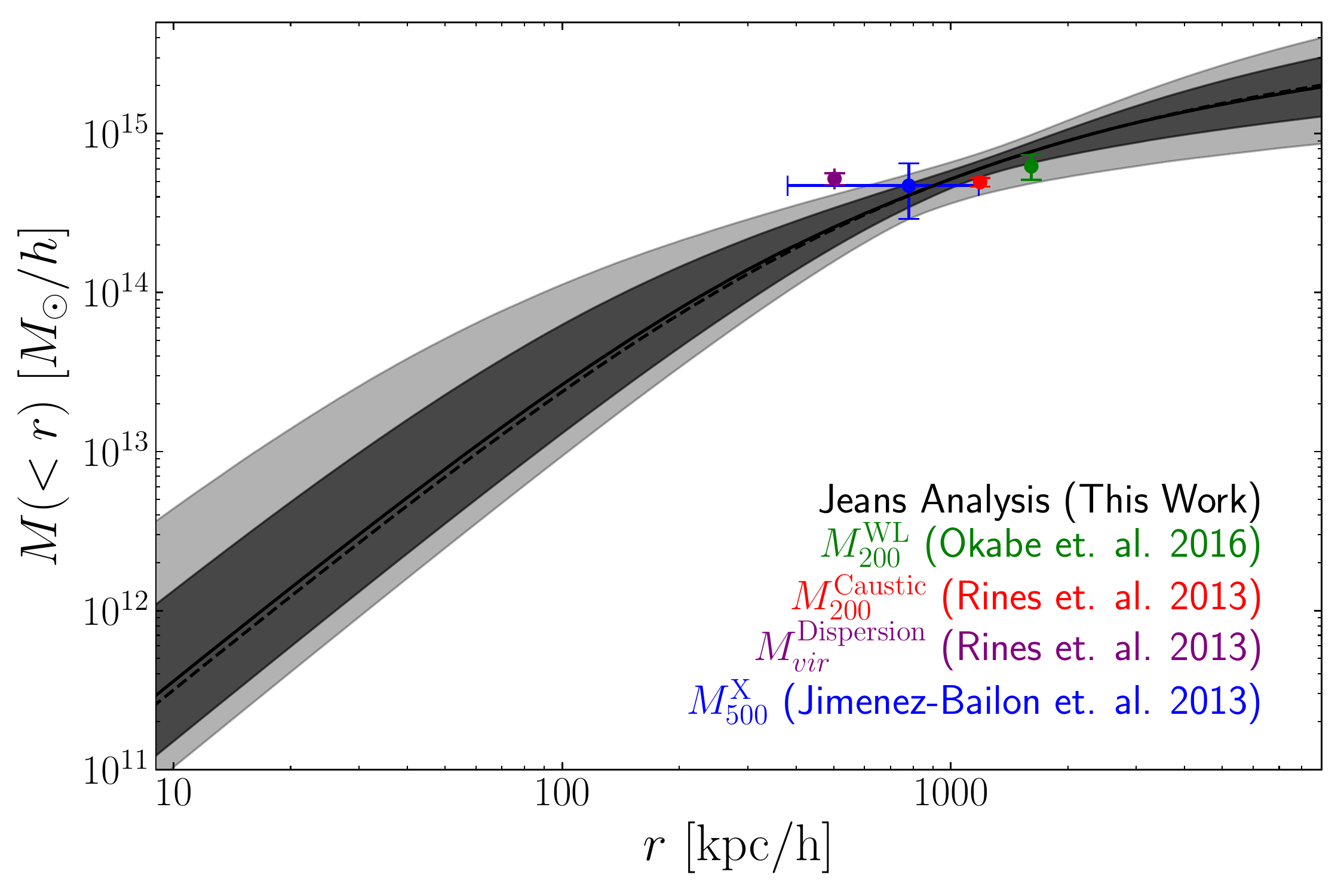}
\caption{Radial mass profile for A267. The darker and lighter regions correspond to $1\sigma$ and $2\sigma$ of the posteriors, respectively. The solid black line is the median posterior curve, while the dotted black line corresponds to $M(<r)$ for the highest likelihood model. The colored points show 4 different mass estimates of A267 from weak lensing \citep[green,][]{Okabe2016}, caustic \citep[red,][]{Rines2013}, velocity dispersion \citep[purple,][]{Rines2013}, and X-ray \citep[blue,][]{Jimenez2013}.}
\label{MassProf}
\end{figure}

Table \ref{DynamParams2} shows the free parameters used in this model.
Most of these parameters are the same as those used in \S\ref{NoJeans} (see Table \ref{DynamParams}), with the replacement of $\sigma_\mathrm{main}\to\sigma_\mathrm{main}(\mathbf{r})$.
This replaces $\sigma_\mathrm{main}$ with three new parameters: two define the dark matter halo ($M_{200}$ and $c_{200}$) and the anisotropy of the velocity dispersion profile $\beta$.
Furthermore, given the assumptions we make when applying and solving the Jeans Equation, we remove the overall cluster rotation, which reduces the number of free parameters by two.
We discuss in this section the application of an $N_\mathrm{subs}=5$ model, which includes a total of 38 free parameters.

Fig. \ref{MainPDFs} shows the 1D and 2D posterior distributions for the parameters that define the main cluster population.
We are able to reasonably constrain all parameters pertaining to the main cluster halo.
There is a strong $c_{200}-\beta$ degeneracy, and a weaker $M_{200}-c_{200}$ degeneracy.
The $M_{200}-c_{200}$ relation is a usefully cosmological scaling relation that exhibits relatively low scatter.
In Fig. \ref{MassConcentration} we zoom-in on the $M_{200}-c_{200}$ panel and compare the results from three models ($N_\mathrm{subs}=0,3,5$) with predictions of the $M_{200}-c_{200}$ relation derived from N-body simulations \citep[][]{Dutton2014}.
Even though the mass of the dark matter halo decreases with increasing number of subpopulations and the posterior for $c_{200}$ is relatively less constrained, we still recover a mean concentration in accordance with \citet{Dutton2014}.

Fig. \ref{DispProf} shows the velocity dispersion profile inferred for A267 in black.
The dark and light shaded regions show the 68\% and 95\% confidence intervals of the posterior PDFs.
The solid and dashed black curves show the median posterior and maximum likelihood velocity dispersion profiles, respectively.
We compare this curve to the velocity dispersion profile calculated by binning the data and weighting these samples by their probability of membership to the main cluster halo which is shown in blue.
In order to qualitatively show any potential bias in calculating a binned velocity dispersion this way, we show in red a posterior predictive check using simulated data.
For each point in the posterior we generate a sample of 1000 member galaxies to the cluster.
We then bin each sample and calculate the velocity dispersion of the binned data in the same way as the blue points.
Shown in red is the composite of these binned samples over the entire posterior.
The posterior predictive check shows that our true galaxy sample (blue points) is consistent with the simulated data (red points).

Fig. \ref{MassProf} shows the mass profile of the dark matter halo for A267.
Like previous plots, the dark and light regions show the 68\% and 95\% confidence intervals of the PDFs, and the solid and dashed black curves show the median posterior and maximum likelihood velocity dispersion profiles, respectively.
For comparison we include previous mass measurements of A267 from a variety of different techniques: in green the weak lensing mass $M^\mathrm{WL}_\mathrm{200}$ \citep[][]{Okabe2016}, in red and purple the caustic mass $M^\mathrm{Caustic}_{200}$ and viral mass calculated with velocity dispersion of cluster members $M^\mathrm{Dispersion}_{vir}$, respectively \citep[][]{Rines2013}, and in blue we show the X-ray derived mass $M^X_{500}$ \citep[][]{Jimenez2013}.
Our results are consistent with $M^X_{500}$, $M^\mathrm{Caustic}_{200}$, and $M^\mathrm{WL}_\mathrm{200}$, but we measure a significantly smaller mass than $M^\mathrm{Dispersion}_{vir}$ because the mass estimate derived from the velocity dispersion is more susceptible to substructure.

\section{Conclusions}
\label{conclusions}

We have developed a muli-population mixture model in order to simultaneously model the internal kinematics and substructure of A267.
We included in this model the ability to fit $N_\mathrm{subs}$ subpopulations, as well as cluster parameters such as NFW scale radius, mean cluster redshift, velocity dispersion, and overall cluster rotation.
We embedded this model in a full Bayesian framework, such that, we quantify posteriors of all free parameters as well as parameter covariances.
In the application of this model to A267, we considered two alternative models that differ in how the cluster velocity dispersion is treated.
We first assumed a simple uniform velocity dispersion profile to analyze the dependence of the internal kinematics on the arbitrary choice of the number of subpopulations.
We then solved the spherical Jeans Equation in order to fit a dark matter halo to A267, thus inferring the enclosed mass profile while allowing the velocity dispersion to vary with radius.

In our first analyses, we investigated the dependence of the internal kinematics on the number of subpopulations.
We showed that as the number of subpopulations increases, the inferred scale radius and velocity dispersion of the cluster both decrease, with significant consequences for cluster mass estimates.
For A267, we found that the preferred model, which allowed for up to $N_{\rm subs}=4$ sub-populations, implied a dynamical mass $\sim 20\%$ smaller than the result from a model that neglects sub-populations.
Furthermore, we found that the mean redshift of the cluster is also sensitive to the presence and treatment of subpopulations.
Compared to the no-substructure model, allowing for $N_{\rm subs}=5$ sub-populations lowers our estimate of the cluster mean velocity by $\sim 100$ km s$^{-1}$, approximately doubling the velocity offset between A267 and its BCG.
This demonstrates how accounting for substructure can have significant implications for detecting ``wobble'' of the BCG around the cluster core, as predicted by self-interacting dark matter models \citep[][]{Harvey2017, Kim2017}.

Finally we embedded our mixture model within a dynamical model that relates the dark matter halo potential to cluster kinematics.
From this analysis, allowing for up to $N_{\rm subs}=5$ sub-populations, we infer for A267 a halo mass  $M_{200}=6.77\pm1.06\times10^{14}M_\odot/h$ and concentration $\log_{10}c_{200}=0.61\pm0.39$ with velocity dispersion anisotropy $\beta=-0.67\pm1.75$.
The mass and concentration posteriors are consistent with the well established $M_{200}-c_{200}$ relation derived from N-body simulations \citep[][]{Dutton2014}.
The corresponding mass profile (Figure \ref{MassProf}) is in good agreement with previously measured masses of A267 from X-ray and weak-lensing measurements \citep{Jimenez2013,Okabe2010}, as well as the dynamical estimate based on the caustic technique \citep{Rines2013}.
Interestingly, the dynamical mass previously estimated directly from the galaxy velocity dispersion (assuming no sub-substructure; \citealt{Rines2013}) is larger than we infer when we allow for $N_{\rm subs}=5$ sub-populations, but in good agreement with the mass profile we obtain if we restrict our Jeans model to $N_{\rm subs}=0$ sub-populations. 

In summary, we have developed a dynamical mixture model to account for both internal rotation and substructure within galaxy clusters.
Our first application, to Abell 267, illustrates the sensitivity of important dynamical results---mean redshift, scale radius, internal velocity dispersion, and dynamical mass---to the presence and modeling of substructure.
This work adds to mounting evidence that, given the widespread interest in using galaxy clusters for both cosmology and tests of dark matter models, it is necessary to account for such substructure when modeling galaxy kinematic data.
In future work, we will extend this analysis to other galaxy clusters with similarly large and high-quality data sets.  

\section*{Acknowledgements}
We thank Margaret Geller for her comments and suggestions that greatly improved the quality of this paper.
We thank Nobu Okabe for generously providing the weak lensing map.
We also thank Kaustuv Basu and Martin Sommer for reducing the radio observations of A267.
M.G.W. is supported by National Science Foundation grants AST-1313045, AST-1412999 and AST-1813881.
M.M. is supported by NSF grant AST-1312997 and AST-1815403.
E.W.O. is supported by NSF grant AST-1313006 and AST-1815767.  

\bibliography{bib}{}

\begin{thebibliography}{}
\expandafter\ifx\csname natexlab\endcsname\relax\def\natexlab#1{#1}\fi

\bibitem[{{Applegate} {et~al.}(2014){Applegate}, {von der Linden}, {Kelly},
  {Allen}, {Allen}, {Burchat}, {Burke}, {Ebeling}, {Mantz}, \&
  {Morris}}]{Applegate2014}
{Applegate}, D.~E., {von der Linden}, A., {Kelly}, P.~L., {et~al.} 2014,
  \mnras, 439, 48

\bibitem[{{Aryal} {et~al.}(2013){Aryal}, {Bhattarai}, {Dhakal}, {Rajbahak}, \&
  {Saurer}}]{Aryal2013}
{Aryal}, B., {Bhattarai}, H., {Dhakal}, S., {Rajbahak}, C., \& {Saurer}, W.
  2013, \mnras, 434, 1939

\bibitem[{{Barreira} {et~al.}(2015){Barreira}, {Li}, {Jennings}, {Merten},
  {King}, {Baugh}, \& {Pascoli}}]{Barreira2015}
{Barreira}, A., {Li}, B., {Jennings}, E., {et~al.} 2015, \mnras, 454, 4085

\bibitem[{{Barrena} {et~al.}(2007){Barrena}, {Boschin}, {Girardi}, \&
  {Spolaor}}]{Barrena2007}
{Barrena}, R., {Boschin}, W., {Girardi}, M., \& {Spolaor}, M. 2007, \aap, 469,
  861

\bibitem[{{Beers} {et~al.}(1982){Beers}, {Geller}, \& {Huchra}}]{Beers1982}
{Beers}, T.~C., {Geller}, M.~J., \& {Huchra}, J.~P. 1982, \apj, 257, 23

\bibitem[{{Bianconi} {et~al.}(2013){Bianconi}, {Ettori}, \&
  {Nipoti}}]{Bianconi2013}
{Bianconi}, M., {Ettori}, S., \& {Nipoti}, C. 2013, \mnras, 434, 1565

\bibitem[{{Binney} \& {Tremaine}(2008)}]{Binney2008}
{Binney}, J., \& {Tremaine}, S. 2008, {Galactic Dynamics: Second Edition}
  (Princeton University Press)

\bibitem[{{Biviano} {et~al.}(2002){Biviano}, {Katgert}, {Thomas}, \&
  {Adami}}]{Biviano2002}
{Biviano}, A., {Katgert}, P., {Thomas}, T., \& {Adami}, C. 2002, \aap, 387, 8

\bibitem[{{Biviano} {et~al.}(2016){Biviano}, {van der Burg}, {Muzzin},
  {Sartoris}, {Wilson}, \& {Yee}}]{Biviano2016}
{Biviano}, A., {van der Burg}, R.~F.~J., {Muzzin}, A., {et~al.} 2016, \aap,
  594, A51

\bibitem[{{Blanton} {et~al.}(2017){Blanton}, {Bershady}, {Abolfathi},
  {Albareti}, {Allende Prieto}, {Almeida}, {Alonso-Garc{\'{\i}}a}, {Anders},
  {Anderson}, {Andrews}, \& et~al.}]{SDSSDR14}
{Blanton}, M.~R., {Bershady}, M.~A., {Abolfathi}, B., {et~al.} 2017, \aj, 154,
  28

\bibitem[{{Chon} {et~al.}(2012){Chon}, {B{\"o}hringer}, \& {Smith}}]{Chon2012}
{Chon}, G., {B{\"o}hringer}, H., \& {Smith}, G.~P. 2012, \aap, 548, A59

\bibitem[{{Churazov} {et~al.}(2015){Churazov}, {Vikhlinin}, \&
  {Sunyaev}}]{Churazov2015}
{Churazov}, E., {Vikhlinin}, A., \& {Sunyaev}, R. 2015, \mnras, 450, 1984

\bibitem[{{Coziol} {et~al.}(2009){Coziol}, {Andernach}, {Caretta},
  {Alamo-Mart{\'{\i}}nez}, \& {Tago}}]{Coziol2009}
{Coziol}, R., {Andernach}, H., {Caretta}, C.~A., {Alamo-Mart{\'{\i}}nez},
  K.~A., \& {Tago}, E. 2009, \aj, 137, 4795

\bibitem[{{Diaferio}(1999)}]{Diaferio1999}
{Diaferio}, A. 1999, \mnras, 309, 610

\bibitem[{{Diaferio} \& {Geller}(1997)}]{Diaferio1997}
{Diaferio}, A., \& {Geller}, M.~J. 1997, \apj, 481, 633

\bibitem[{{Dressler} \& {Shectman}(1988)}]{Dressler1988}
{Dressler}, A., \& {Shectman}, S.~A. 1988, \aj, 95, 985

\bibitem[{{Dubinski} \& {Carlberg}(1991)}]{Dubinski1991}
{Dubinski}, J., \& {Carlberg}, R.~G. 1991, \apj, 378, 496

\bibitem[{{Dutton} \& {Macci{\`o}}(2014)}]{Dutton2014}
{Dutton}, A.~A., \& {Macci{\`o}}, A.~V. 2014, \mnras, 441, 3359

\bibitem[{{Einasto} {et~al.}(2012){Einasto}, {Vennik}, {Nurmi}, {Tempel},
  {Ahvensalmi}, {Tago}, {Liivam{\"a}gi}, {Saar}, {Hein{\"a}m{\"a}ki},
  {Einasto}, \& {Mart{\'{\i}}nez}}]{Einasto2012}
{Einasto}, M., {Vennik}, J., {Nurmi}, P., {et~al.} 2012, \aap, 540, A123

\bibitem[{{Feroz} {et~al.}(2009){Feroz}, {Hobson}, \& {Bridges}}]{Feroz2009}
{Feroz}, F., {Hobson}, M.~P., \& {Bridges}, M. 2009, \mnras, 398, 1601

\bibitem[{{Geller}(1984)}]{Geller1984}
{Geller}, M.~J. 1984, Comments on Astrophysics, 10, 47

\bibitem[{{Geller} {et~al.}(2013){Geller}, {Diaferio}, {Rines}, \&
  {Serra}}]{Geller2013}
{Geller}, M.~J., {Diaferio}, A., {Rines}, K.~J., \& {Serra}, A.~L. 2013, \apj,
  764, 58

\bibitem[{{Geller} {et~al.}(2014){Geller}, {Hwang}, {Diaferio}, {Kurtz}, {Coe},
  \& {Rines}}]{Geller2014}
{Geller}, M.~J., {Hwang}, H.~S., {Diaferio}, A., {et~al.} 2014, \apj, 783, 52

\bibitem[{{Gifford} {et~al.}(2013){Gifford}, {Miller}, \& {Kern}}]{Gifford2013}
{Gifford}, D., {Miller}, C., \& {Kern}, N. 2013, \apj, 773, 116

\bibitem[{{Girardi} {et~al.}(2015){Girardi}, {Mercurio}, {Balestra}, {Nonino},
  {Biviano}, {Grillo}, {Rosati}, {Annunziatella}, {Demarco}, {Fritz}, {Gobat},
  {Lemze}, {Presotto}, {Scodeggio}, {Tozzi}, {Bartosch Caminha}, {Brescia},
  {Coe}, {Kelson}, {Koekemoer}, {Lombardi}, {Medezinski}, {Postman},
  {Sartoris}, {Umetsu}, {Zitrin}, {Boschin}, {Czoske}, {De Lucia}, {Kuchner},
  {Maier}, {Meneghetti}, {Monaco}, {Monna}, {Munari}, {Seitz}, {Verdugo}, \&
  {Ziegler}}]{Girardi2015}
{Girardi}, M., {Mercurio}, A., {Balestra}, I., {et~al.} 2015, \aap, 579, A4

\bibitem[{{Girardi} {et~al.}(2016){Girardi}, {Boschin}, {Gastaldello},
  {Giovannini}, {Govoni}, {Murgia}, {Barrena}, {Ettori}, {Trasatti}, \&
  {Vacca}}]{Girardi2016}
{Girardi}, M., {Boschin}, W., {Gastaldello}, F., {et~al.} 2016, \mnras, 456,
  2829

\bibitem[{{Gonzalez} {et~al.}(2015){Gonzalez}, {Fo{\"e}x}, {Nilo
  Castell{\'o}n}, {Dom{\'{\i}}nguez Romero}, {Alonso}, {Garc{\'{\i}}a Lambas},
  {Moreschi}, \& {Gallo}}]{Gonzalez2015}
{Gonzalez}, E.~J., {Fo{\"e}x}, G., {Nilo Castell{\'o}n}, J.~L., {et~al.} 2015,
  \mnras, 452, 2225

\bibitem[{{Guennou} {et~al.}(2014){Guennou}, {Biviano}, {Adami}, {Limousin},
  {Lima Neto}, {Mamon}, {Ulmer}, {Gavazzi}, {Cypriano}, {Durret}, {Clowe},
  {LeBrun}, {Allam}, {Basa}, {Benoist}, {Cappi}, {Halliday}, {Ilbert},
  {Johnston}, {Jullo}, {Just}, {Kubo}, {M{\'a}rquez}, {Marshall}, {Martinet},
  {Maurogordato}, {Mazure}, {Murphy}, {Plana}, {Rostagni}, {Russeil},
  {Schirmer}, {Schrabback}, {Slezak}, {Tucker}, {Zaritsky}, \&
  {Ziegler}}]{Guennou2014}
{Guennou}, L., {Biviano}, A., {Adami}, C., {et~al.} 2014, \aap, 566, A149

\bibitem[{{Harvey} {et~al.}(2017){Harvey}, {Courbin}, {Kneib}, \&
  {McCarthy}}]{Harvey2017}
{Harvey}, D., {Courbin}, F., {Kneib}, J.~P., \& {McCarthy}, I.~G. 2017, \mnras,
  472, 1972

\bibitem[{{Hou} {et~al.}(2009){Hou}, {Parker}, {Harris}, \& {Wilman}}]{Hou2009}
{Hou}, A., {Parker}, L.~C., {Harris}, W.~E., \& {Wilman}, D.~J. 2009, \apj,
  702, 1199

\bibitem[{{Hwang} \& {Lee}(2007)}]{Hwang2007}
{Hwang}, H.~S., \& {Lee}, M.~G. 2007, \apj, 662, 236

\bibitem[{{Jim{\'e}nez-Bail{\'o}n} {et~al.}(2013){Jim{\'e}nez-Bail{\'o}n},
  {Lozada-Mu{\~n}oz}, \& {Aguerri}}]{Jimenez2013}
{Jim{\'e}nez-Bail{\'o}n}, E., {Lozada-Mu{\~n}oz}, M., \& {Aguerri}, J.~A.~L.
  2013, Astronomische Nachrichten, 334, 377

\bibitem[{{Jones} {et~al.}(2009){Jones}, {Nulsen}, {Arnaud}, {Donahue},
  {Mahdavi}, {Vrtilek}, {Nakazawa}, {Randall}, \& {Ruszkowski}}]{Jones2009}
{Jones}, C., {Nulsen}, P., {Arnaud}, K., {et~al.} 2009, in Bulletin of the
  American Astronomical Society, Vol.~41, American Astronomical Society Meeting
  Abstracts \#213, 351

\bibitem[{{Kalinkov} {et~al.}(2005){Kalinkov}, {Valchanov}, {Valtchanov},
  {Kuneva}, \& {Dissanska}}]{Kalinkov2005}
{Kalinkov}, M., {Valchanov}, T., {Valtchanov}, I., {Kuneva}, I., \&
  {Dissanska}, M. 2005, \mnras, 359, 1491

\bibitem[{Kass \& Raftery(1995)}]{Kass1995}
Kass, R.~E., \& Raftery, A.~E. 1995, Journal of the American Statistical
  Association, 90, 773

\bibitem[{{Kim} {et~al.}(2017){Kim}, {Peter}, \& {Wittman}}]{Kim2017}
{Kim}, S.~Y., {Peter}, A.~H.~G., \& {Wittman}, D. 2017, \mnras, 469, 1414

\bibitem[{{Kneib}(2008)}]{Kneib2008}
{Kneib}, J.-P. 2008, in Lecture Notes in Physics, Berlin Springer Verlag, Vol.
  740, A Pan-Chromatic View of Clusters of Galaxies and the Large-Scale
  Structure, ed. M.~{Plionis}, O.~{L{\'o}pez-Cruz}, \& D.~{Hughes}, 24

\bibitem[{{Li}(1998)}]{Li1998}
{Li}, L.-X. 1998, General Relativity and Gravitation, 30, 497

\bibitem[{{Mamon} {et~al.}(2013){Mamon}, {Biviano}, \& {Bou{\'e}}}]{Mamon2013}
{Mamon}, G.~A., {Biviano}, A., \& {Bou{\'e}}, G. 2013, \mnras, 429, 3079

\bibitem[{{Mamon} \& {{\L}okas}(2005)}]{Mamon2005}
{Mamon}, G.~A., \& {{\L}okas}, E.~L. 2005, \mnras, 363, 705

\bibitem[{{Manolopoulou} \& {Plionis}(2017)}]{Manolopoulou2017}
{Manolopoulou}, M., \& {Plionis}, M. 2017, \mnras, 465, 2616

\bibitem[{{Materne} \& {Hopp}(1983)}]{Materne1983}
{Materne}, J., \& {Hopp}, U. 1983, \aap, 124, L13

\bibitem[{{Moffat} \& {Rahvar}(2014)}]{Moffat2014}
{Moffat}, J.~W., \& {Rahvar}, S. 2014, \mnras, 441, 3724

\bibitem[{{Navarro} {et~al.}(1996){Navarro}, {Frenk}, \& {White}}]{Navarro1996}
{Navarro}, J.~F., {Frenk}, C.~S., \& {White}, S.~D.~M. 1996, \apj, 462, 563

\bibitem[{{Navarro} {et~al.}(1997){Navarro}, {Frenk}, \& {White}}]{Navarro1997}
---. 1997, \apj, 490, 493

\bibitem[{{Oegerle} \& {Hill}(1992)}]{Oegerle1992}
{Oegerle}, W.~R., \& {Hill}, J.~M. 1992, \aj, 104, 2078

\bibitem[{{Okabe} \& {Smith}(2016)}]{Okabe2016}
{Okabe}, N., \& {Smith}, G.~P. 2016, \mnras, 461, 3794

\bibitem[{{Okabe} {et~al.}(2010){Okabe}, {Takada}, {Umetsu}, {Futamase}, \&
  {Smith}}]{Okabe2010}
{Okabe}, N., {Takada}, M., {Umetsu}, K., {Futamase}, T., \& {Smith}, G.~P.
  2010, \pasj, 62, 811

\bibitem[{{Old} {et~al.}(2017){Old}, {Wojtak}, {Pearce}, {Gray}, {Mamon},
  {Sif{\'o}n}, {Tempel}, {Biviano}, {Yee}, {de Carvalho}, {M{\"u}ller}, {Sepp},
  {Skibba}, {Croton}, {Power}, {von der Linden}, \& {Saro}}]{Old2017}
{Old}, L., {Wojtak}, R., {Pearce}, F.~R., {et~al.} 2017, ArXiv e-prints,
  arXiv:1709.10108

\bibitem[{{Pinkney} {et~al.}(1996){Pinkney}, {Roettiger}, {Burns}, \&
  {Bird}}]{Pinkney1996}
{Pinkney}, J., {Roettiger}, K., {Burns}, J.~O., \& {Bird}, C.~M. 1996, \apjs,
  104, 1

\bibitem[{{Postman} {et~al.}(2012){Postman}, {Coe}, {Ben{\'{\i}}tez},
  {Bradley}, {Broadhurst}, {Donahue}, {Ford}, {Graur}, {Graves}, {Jouvel},
  {Koekemoer}, {Lemze}, {Medezinski}, {Molino}, {Moustakas}, {Ogaz}, {Riess},
  {Rodney}, {Rosati}, {Umetsu}, {Zheng}, {Zitrin}, {Bartelmann}, {Bouwens},
  {Czakon}, {Golwala}, {Host}, {Infante}, {Jha}, {Jimenez-Teja}, {Kelson},
  {Lahav}, {Lazkoz}, {Maoz}, {McCully}, {Melchior}, {Meneghetti}, {Merten},
  {Moustakas}, {Nonino}, {Patel}, {Reg{\"o}s}, {Sayers}, {Seitz}, \& {Van der
  Wel}}]{Postman2012}
{Postman}, M., {Coe}, D., {Ben{\'{\i}}tez}, N., {et~al.} 2012, \apjs, 199, 25

\bibitem[{{Rabitz} {et~al.}(2017){Rabitz}, {Zhang}, {Schwope}, {Verdugo},
  {Reiprich}, \& {Klein}}]{Rabitz2017}
{Rabitz}, A., {Zhang}, Y.-Y., {Schwope}, A., {et~al.} 2017, \aap, 597, A24

\bibitem[{{Regos} \& {Geller}(1989)}]{Regos1989}
{Regos}, E., \& {Geller}, M.~J. 1989, \aj, 98, 755

\bibitem[{{Richardson} {et~al.}(2011){Richardson}, {Irwin}, {McConnachie},
  {Martin}, {Dotter}, {Ferguson}, {Ibata}, {Chapman}, {Lewis}, {Tanvir}, \&
  {Rich}}]{Richardson2011}
{Richardson}, J.~C., {Irwin}, M.~J., {McConnachie}, A.~W., {et~al.} 2011, \apj,
  732, 76

\bibitem[{{Rines} {et~al.}(2013){Rines}, {Geller}, {Diaferio}, \&
  {Kurtz}}]{Rines2013}
{Rines}, K., {Geller}, M.~J., {Diaferio}, A., \& {Kurtz}, M.~J. 2013, \apj,
  767, 15

\bibitem[{{Rines} {et~al.}(2003){Rines}, {Geller}, {Kurtz}, \&
  {Diaferio}}]{Rines2003}
{Rines}, K., {Geller}, M.~J., {Kurtz}, M.~J., \& {Diaferio}, A. 2003, \aj, 126,
  2152

\bibitem[{{Rines} {et~al.}(2016){Rines}, {Geller}, {Diaferio}, \&
  {Hwang}}]{Rines2016}
{Rines}, K.~J., {Geller}, M.~J., {Diaferio}, A., \& {Hwang}, H.~S. 2016, \apj,
  819, 63

\bibitem[{{Sohn} {et~al.}(2017){Sohn}, {Geller}, {Zahid}, {Fabricant},
  {Diaferio}, \& {Rines}}]{Sohn2017}
{Sohn}, J., {Geller}, M.~J., {Zahid}, H.~J., {et~al.} 2017, \apjs, 229, 20

\bibitem[{{Stock} {et~al.}(2015){Stock}, {Meyer}, {Sarli}, {Bartelmann},
  {Balestra}, {Grillo}, {Koekemoer}, {Mercurio}, {Nonino}, \&
  {Rosati}}]{Stock2015}
{Stock}, D., {Meyer}, S., {Sarli}, E., {et~al.} 2015, \aap, 584, A63

\bibitem[{{Sunyaev} \& {Zeldovich}(1970)}]{Sunyaev1970}
{Sunyaev}, R.~A., \& {Zeldovich}, Y.~B. 1970, \apss, 9, 368

\bibitem[{{Tasca} {et~al.}(2016){Tasca}, {Le Fevre}, {Ribeiro}, {Thomas},
  {Moreau}, {Cassata}, {Garilli}, {Le Brun}, {Lemaux}, {Maccagni},
  {Pentericci}, {Schaerer}, {Vanzella}, {Zamorani}, {Zucca}, {Amorin},
  {Bardelli}, {Cassara}, {Castellano}, {Cimatti}, {Cucciati}, {Durkalec},
  {Fontana}, {Giavalisco}, {Grazian}, {Hathi}, {Ilbert}, {Paltani}, {Pforr},
  {Scodeggio}, {Sommariva}, {Talia}, {Tresse}, {Vergani}, {Capak}, {Charlot},
  {Contini}, {de la Torre}, {Dunlop}, {Fotopoulou}, {Guaita}, {Koekemoer},
  {Lopez-Sanjuan}, {Mellier}, {Salvato}, {Scoville}, {Taniguchi}, \&
  {Wang}}]{Tasca2016}
{Tasca}, L.~A.~M., {Le Fevre}, O., {Ribeiro}, B., {et~al.} 2016, ArXiv
  e-prints, arXiv:1602.01842

\bibitem[{{Tovmassian}(2015)}]{Tovmassian2015}
{Tovmassian}, H.~M. 2015, Astrophysics, 58, 328

\bibitem[{{Tucker} {et~al.}(2017){Tucker}, {Walker}, {Mateo}, {Olszewski},
  {Bailey}, {Crane}, \& {Shectman}}]{Tucker2017}
{Tucker}, E., {Walker}, M.~G., {Mateo}, M., {et~al.} 2017, \aj, 154, 113

\bibitem[{{van Haarlem} {et~al.}(1993){van Haarlem}, {Cayon}, {Gutierrez de La
  Cruz}, {Martinez-Gonzalez}, \& {Rebolo}}]{vanHaarlem1993}
{van Haarlem}, M.~P., {Cayon}, L., {Gutierrez de La Cruz}, C.,
  {Martinez-Gonzalez}, E., \& {Rebolo}, R. 1993, \mnras, 264, 71

\bibitem[{{Vikhlinin} {et~al.}(2009){Vikhlinin}, {Kravtsov}, {Burenin},
  {Ebeling}, {Forman}, {Hornstrup}, {Jones}, {Murray}, {Nagai}, {Quintana}, \&
  {Voevodkin}}]{Vikhlinin2009}
{Vikhlinin}, A., {Kravtsov}, A.~V., {Burenin}, R.~A., {et~al.} 2009, \apj, 692,
  1060

\bibitem[{{Voit}(2005)}]{Voit2005}
{Voit}, G.~M. 2005, Reviews of Modern Physics, 77, 207

\bibitem[{{Walker} {et~al.}(2009){Walker}, {Mateo}, {Olszewski}, {Sen}, \&
  {Woodroofe}}]{Walker2009}
{Walker}, M.~G., {Mateo}, M., {Olszewski}, E.~W., {Sen}, B., \& {Woodroofe}, M.
  2009, \aj, 137, 3109

\bibitem[{{West} \& {Bothun}(1990)}]{West1990}
{West}, M.~J., \& {Bothun}, G.~D. 1990, \apj, 350, 36

\bibitem[{{West} {et~al.}(1988){West}, {Oemler}, \& {Dekel}}]{West1988}
{West}, M.~J., {Oemler}, Jr., A., \& {Dekel}, A. 1988, \apj, 327, 1

\bibitem[{{Wojtak} {et~al.}(2013){Wojtak}, {Gottl{\"o}ber}, \&
  {Klypin}}]{Wojtak2013}
{Wojtak}, R., {Gottl{\"o}ber}, S., \& {Klypin}, A. 2013, \mnras, 434, 1576

\bibitem[{{Zhang} {et~al.}(2012){Zhang}, {Verdugo}, {Klein}, \&
  {Schneider}}]{Zhang2012}
{Zhang}, Y.-Y., {Verdugo}, M., {Klein}, M., \& {Schneider}, P. 2012, \aap, 542,
  A106

\end{thebibliography}
\bibliographystyle{aasjournal}
\label{lastpage}
\end{document}